\documentclass[aps,prb,reprint,groupedaddress]{revtex4-1}

\usepackage{graphics}
\usepackage{amsmath}
\usepackage{notes2bib}
\begin{document}


\title{Renormalization group analysis of the finite 2DXY model with fourfold anisotropy: Application to the magnetic susceptibility of a ferromagnetic ultrathin film}


\author{D. Venus}
\email{[corresponding author] venus@physics.mcmaster.ca}
\affiliation{Department of Physics and Astronomy, McMaster University, Hamilton, Ontario, Canada}

\date{\today}

\begin{abstract}
The renormalization group equations describing a finite  2DXY system with fourfold anisotropy are solved in two steps, in order to study the magnetic transition to paramagnetism in an ultrathin film.  First, the equations are linearized near the critical coupling $K=J/k_BT=2/\pi$.  This allows integration constants to be evaluated at the fixed point, and the tuning of the constants to represent a ferromagnetic ultrathin film. An exact solution of the linearized equations confirms that a finite-size Kosterlitz-Thouless (KT) transition occurs in the presence of weak fourfold anisotropy, and that an Ising transition occurs for strong anisotropy.  The behaviour of a given system, and the crossover region between these two types of transitions, is determined by the system parameters through the product of an anisotropy parameter and the logarithm of the system size.  The linearized RG equations are not quantitatively reliable across the extended temperature range of the finite-size transition, but they do define the parameter space where a second approach, where the fourfold anisotropy is treated as a perturbation, is valid.  This perturbative treatment provides a quantitative determination of the renormalized exchange coupling, vortex density, and anisotropy throughout the transition. In particular, the coupling has a universal point of inflection where vortex-antivortex pairs unbind (as opposed to a ``universal jump"), and goes to zero asymptotically in the paramagnetic state, as is expected for a finite system.  These results are used to calculate the magnetic susceptibility as the system moves from one dominated by spin waves to one dominated by a free vortex gas.   The presence of anisotropy makes it necessary to include both the susceptibility $\chi_{||}$ due to fluctuations of the magnitude of the magnetization, and $\chi_\perp$ due to angular fluctuations of the magnetization about a fourfold easy axis.  A comparison to recent measurements of the magnetic susceptibility of ultrathin Fe/W(001) films suggests that a detailed quantitative analysis of the experimental results can provide information on vortex formation, the disappearance of anisotropy, and dissipative processes in the finite-size KT transition of a real system.
\end{abstract}


\maketitle

\section{Introduction}

Nearly half a century after Berezinskii\cite{Berezinskii}, Kosterlitz  and Thouless\cite{Kosterlitz1}, and Kosterlitz \cite{Kosterlitz2}, introduced the ideas underlying the transition of a spin system between phases supporting excitations of different topologies, there has been a resurgence of interest in topological spin textures in material systems.  Topological spin textures of current interest include chiral ``bubbles" in perpendicularly magnetized films\cite{Jenkins},  vortices bounded within a ferromagnetic microstructure\cite{Sugimoto,Kuepper}, 2D skyrmions\cite{Yu} in ferromagnetic and antiferromagnetic layers, 3D skyrmions in crystals\cite{Muhlbauer,Munzer}, as well as the chiral domain wall spin textures in these systems\cite{Clarke,Gungordu}.  There is great interest in the phase diagrams and transitions between the topological phases\cite{Rybakov}, as well as in the non-equilibrium dynamics of the topological excitations\cite{Fu}.  These investigations are driven both by fundamental questions in the larger area of topological materials, and by the prospect of potential applications in spintronics.

Within the context of this larger field of work, the simpler, archetypical 2DXY ferromagnetic system originally considered by Kosterlitz and Thouless\cite{Kosterlitz1,Kosterlitz2,Kosterlitz3} can provide insight into basic questions relevant to many of the systems that support topological spin textures.  These include the effects of dissipation and relaxation towards equilibrium\cite{He1,Ozeki,Rojas}, finite-size effects\cite{Bramwell1,Chung}, fluctuations near transitions\cite{Archambault2} and the influence of perturbations from an ideal symmetry as may be provided by, for example, defects\cite{Holl} and anisotropies\cite{Jose2}. 

Despite these opportunities, there is a relatively small literature reporting experimental studies of  2DXY ferromagnetic films focused on the Kosterlitz-Thouless  (KT) transition and vortex dynamics. Experimental work has focused instead on superconductors\cite{Goldman} and layered three dimensional antiferromagnets\cite{Nielson,Bramwell4,Cuccoli}. For the most part, studies of ferromagnetic films consist of magnetization studies of ferromagnetic films grown on (001)-oriented metallic substrates\cite{Taroni}, where a non-Ising magnetization exponent was reported and only later interpreted as the signature of a finite-size KT transition\cite{Bramwell1}.  Early indications that experimental measurements of the magnetic susceptibility displayed the characteristic form of a KT transition in such films\cite{Elmers1} were made quantitative only recently\cite{Atchison}, using Fe/W(001) ultrathin films.  This opens the path for further quantitative experimental studies using the magnetic susceptibility, aimed at a better understanding of topological spin textures using this simple realization of a topological phase transition.

The first step is to make contact between the experimental thin film results and theoretical predictions through a quantitative determination of the exchange and anisotropy parameters in a 2DXY system.  Kosterlitz and Thouless\cite{Kosterlitz1,Kosterlitz2} treated the \textit{infinite, isotropic} system and introduced the KT transition.  Bramwell, Holdsworth and coworkers\cite{Bramwell1,Bramwell2,Archambault2} investigated the \textit{finite, isotropic} system and identified essential finite-size effects and the finite-size KT transition with separate characteristic temperatures for the formation of vortex-antivortex pairs and for unbinding of the pairs to form a free vortex gas.  Jos{\'{e}} {\it{et al.}}\cite{Jose2,Jose} derived the renormalization group equations for the \textit{infinite, anisotropic} system with an $n$-fold, in-plane anisotropy.  They showed that the system is Ising-like for $n < 4$, and has a KT transition when $n > 4$.   A system with microscopic fourfold anisotropy flows to a second order transition to paramagnetism with non-universal critical exponents that depend upon the strength of the anisotropy.  For large anisotropy, there is a cross-over to 2D Ising exponents.  

The \textit{finite, anisotropic} 2DXY model with fourfold anisotropy has not been investigated in detail.  Taroni \textit{et al.}\cite{Taroni} have reported Monte Carol simulations as a function of the strength of the fourfold anisotropy, and of system size.  They find a competition between the finite size effects and anisotropy.  For small anisotropy, finite size effects prevail and the value of the effective critical exponent of the magnetization indicates that a finite-size KT transition occurs.   As the anisotropy increases, the effective critical exponent crosses over to the 2D Ising value.  The present article concentrates instead on solutions of the RG equations to provide a detailed description of the behaviour of an ultrathin ferromagnetic film with fourfold anisotropy.  This is accomplished in two stages.  In the first stage, the RG equations are expanded to lowest order about the critical coupling, when $K=J/k_BT = 2/\pi$, and solved exactly.  This allows appropriate physical parameters to be set, and makes contact with the previous findings for the finite, isotropic system. These results validate the use of a second approach within a large parameter space, where the fourfold anisotropy is treated as a perturbation.  This method yields quantitative results for the effective exchange coupling, vortex density and screened fourfold anisotropy as a function of temperature and system size throughout the full temperature range of the finite-size transition. It shows that  the coupling has a universal point of inflection where vortex-antivortex pairs unbind (as opposed to a ``universal jump"), and goes to zero asymptotically in the paramagnetic state.  The fourfold anisotropy also goes to zero smoothly just above the temperature where the vortex pairs unbind.

These quantities are then used to find two components of the magnetic susceptibility:  an improved calculation of the longitudinal susceptibility due to fluctuations in the magnitude of the magnetization\cite{Archambault1}, and, in addition, the transverse susceptibility due to fluctuations of the magnetization direction about an easy axis in the fourfold system.  The appropriate combination of these susceptibilities give an excellent qualitative account of the experimental measurements of Atchison \textit{et al.}\cite{Atchison} and suggest that a detailed quantitative comparison with  the measurements can provide information on vortex formation, the disappearance of anisotropy, and domain processes in the finite-size KT transition of an experimental system.


%

\section{Renormalization group equations under geometric scaling}

The 2DXY ferromagnetic spin model with a fourfold anisotropy is represented by the Hamiltonian\cite{Jose2}
\begin{equation}
\label{Ham}
H=J \sum_{i,j} [1 - \cos(\theta_i - \theta_j )] + \sum_i h_4 [1- \cos(4\theta_i)],
\end{equation}
where $J$ is the bare nearest neighbour exchange coupling, $\theta_i$ is the angle the in-plane spin at lattice site $i$ makes with a fixed in-plane ``easy" magnetic axis, and $h_4$ is the microscopic anisotropy energy for a fourfold in-plane anisotropy.  The sum over $i,j$ is over nearest neighbours on a square lattice.  The renormalization group (RG) equations under geometric scaling for this model  have been derived by Jos{\'{e}} {\it{et al.}}\cite{Jose,Jose2} using an approximation due to Villain\cite{Villain} that is applicable for small anisotropy. After defining $K=J/k_B T$ as the temperature-normalized exchange coupling, they find:
\begin{equation}
\label{RG1}
\frac{dK^{-1}}{d\ell} = 4\pi^3 y_0^2 e^{-\pi^2 K} - 16\pi y_4^2 K^{-2} e^{-4K^{-1}},
\end{equation}
\begin{equation}
\label{RG2}
\frac{dy_0}{d\ell} = (2-\pi K) y_0,
\end{equation}
\begin{equation}
\label{RG3}
\frac{dy_4}{d\ell} = (2-\frac{4}{\pi} K^{-1}) y_4.
\end{equation}
These equations are first order in the system variables $y_0$ and $y_4$ (see below), with corrections in the third order.  $\ell = \ln L$ is the scaling length, where the system size $L$ is in units of the nearest neighbour lattice constant.  The temperature-normalized anisotropy is
\begin{equation}
y_4 = \frac{h_4}{2k_BT}.
\label{y4}
\end{equation}
Due to approximations made in the Villain model, the maximum value of $y_4 =1$ corresponds to a clock model with effectively infinite anisotropy.\cite{Jose2}

The fugacity of a vortex, or the  density of vortices, is given by
\begin{equation}
Y =y_0 e^{\frac{-\pi^2 K}{2}}.
\label{Ydef}
\end{equation}
$y_0$ is a small parameter that is renormalized by the flow described by the RG equations.  It has been introduced into the Villain model (where $y_0=1$), to create the generalized Villain model. In the regions $K>2/\pi$, $y_0$ is renormalized to zero and is an irrelevant parameter.  In the region $K<2/\pi$, $y_0$ is renormalized to unity and is a relevant parameter.  In this way, it can be shown\cite{Jose2} that the generalized Villain model formally reproduces the results for an isotropic 2DXY system as described by the Villain model.  By introducing $y_0$, it was possible to display a dual symmetry between $y_0$ and $y_4$ in the anisotropic 2DXY model.  This was instrumental in the original derivation of the RG equations.

While introducing $y_0$ was necessary, it is problematic when relating the system of equations to a physical system.  This is because $y_0$ is a phenomenological, and not a physical, parameter.  For this reason, it is useful to recast the equations in terms of the fugacity $Y$, and a modified, but still physical, anisotropy parameter $Y_4$ that maintains the dual relation with $Y$:
\begin{equation}
Y_4=y_4 e^{-2K^{-1}}.
\label{Y4def}
\end{equation}

A line of fixed points of the RG equations occurs when $y_0 = y_4 = 0$, regardless of the value of $K$. A second line occurs when $y_0 = \pm y_4;\;K=2/\pi$. Jos{\'{e}} {\it{et al.}}\cite{Jose2} have shown that, in an infinite system, this second line of fixed points mark second-order phase transitions for systems of different microscopic fourfold anisotropy.  These transitions have non-universal exponents that depend upon the anisotropy, and, in the limit of strong anisotropy, cross over to 2D Ising transitions.

The RG equations (\ref{RG1}) to (\ref{RG3}) that include fourfold anisotropy will be solved for a finite system in two steps.  In Section III, the equations are solved by expansion about the critical coupling $K=2/\pi$.  This is referred to as the ``critical approximation" and is routinely used to study critical properties. This step is necessary because the fixed point is the only place where boundary conditions are known precisely.  RG equations valid at the fixed points can be used to determine the values of constants of integration, and to tune parameters in the equations to values appropriate for ultrathin films.  The solutions confirm that a finite-size KT transition can be preserved in the presence of fourfold anisotropy, and that, because the fixed point is approached only logarithmically in the system size $L$, there is a large parameter space where the renormalization flow does not approach the critical point closely.  In Section IV, a second approximation is used.  This is referred to as the ``perturbative approximation".  It is not as good close to the fixed point (where the flow does \textit{not} proceed in a finite system), but is valid across the broad temperature range of the finite-size transition where the flow does proceed.  This provides a more accurate calculation of the effective coupling, vortex density, anisotropy, and ultimately, the magnetic susceptibility.

\section{Renormalized coupling near the critical coupling, in the presence of fourfold anisotropy}

\subsection{Flow equations}
This section extends previous work\cite{Bramwell1,Bramwell2} by including the effects of fourfold anisotropy.  The coupling in the RG equations is expanded about its value at the fixed point, to lowest order in the parameter $x$:
\begin{equation}
\label{x}
x=\pi K -2,
\end{equation}
\begin{equation}
Y =y_0 e^{-\pi\frac{x+2}{2}},\:\: Y_4=y_4 e^{-\pi\frac{2}{x+2}}.
\label{Ysdef}
\end{equation}

This yields the equations to lowest order in $x$ for small $x$:
\begin{equation}
\frac{dx}{d\ell} = \gamma^2\: [ -(\frac{x+2}{2})^2Y^2 + Y_4^2] \rightarrow \gamma^2\: ( -Y^2 + Y_4^2),
\label{xofl}
\end{equation}
with $\gamma = 4\pi$,
\begin{equation}
\frac{dY}{d\ell} = -x Y
\label{Yofl}
\end{equation}
\begin{equation}
\frac{dY_4}{d\ell} =  \frac{2}{2+x} \;x\;Y_4 \rightarrow x Y_4.
\label{Y4ofl}
\end{equation}

Using eq.(\ref{Yofl}) and (\ref{Y4ofl}) to substitute for one power of $Y$ and $Y_4$, respectively, eq.(\ref{xofl}) can be written as a perfect differential that can be integrated to give
\begin{equation}
x^2(\ell) =\gamma^2\:[ Y^2(\ell) +Y_4^2(\ell)] +C,
\label{Cdef1}
\end{equation}
where $C$ is a constant of integration.
Taking the ratio of eq.(\ref{Yofl}) and (\ref{Y4ofl}) gives an expression independent of $x$ that can be integrated to yield
\begin{equation}
Y(\ell) Y_4(\ell) = D,
\label{Ddef}
\end{equation}
where $D$ is a second constant of integration.  The known parameter values at the fixed point $x=0\:, Y=\pm Y_4$ require that 
\begin{equation}
Y|_{x=0} = Y_4|_{x=0} = \sqrt{D},
\end{equation}
and indicate that an infinite system with bare anisotropy $h_4$ flows to a fixed point where $Y_4 = \sqrt{D}$.

The constant $C$ is determined by substituting eq.(\ref{Ddef}) into eq.(\ref{Cdef1}) and using the known values at the fixed point, to give $C=-2\gamma^2D$.  
 This value describes a flow line leading to the fixed point.  This value of $C$ therefore defines the separatrix between different types of flow leading away from this fixed point (for a system with this anisotropy). The flow for systems near the separatrix is investigated\cite{Kosterlitz3,Berlinsky} by allowing the integration constant $C$ to deviate from the separatrix by a small amount proportional to the reduced temperature, $t=(T-T_S)/T_S$, where $T_S$ gives flow on the separatrix to the phase transition represented by the fixed point. Then
\begin{equation}
C= -2\gamma^2D -\alpha t,
\label{Cdef}
\end{equation}
 where\cite{Gupta,Bramwell5} $\alpha = (\pi/b)^2$ and $b=1.846...$ It can be shown\footnote{D. Venus, unpublished} that negative values of $t$ correspond to flow away from the fixed point to a low temperature phase with high anisotropy and vanishing vortex density (fugacity).  Positive values of $t$ give flow to a high temperature phase with vanishing anisotropy and high vortex density.  This  type of behaviour led to the original identification of a topological transition at $t=0$ in the isotropic system\cite{Kosterlitz1,Kosterlitz2}.  
 
Finally, the value of $C$ is used to complete the square in eq.(\ref{Cdef1}) to yield an equation that describes the flow near the critical coupling:
\begin{equation}
\begin{split}
&\pm \sqrt{x(\ell)^2 +\alpha t} = \gamma [Y(\ell) - Y_4(\ell)] \\
&=\gamma[Y(\ell) - \frac{D}{Y(\ell)}] =\gamma [\frac{D}{Y_4(\ell)} - Y_4(\ell)].
\end{split}
\label{sep}
\end{equation}

\subsection{Tuning to the physical parameters of the system}
To follow the renormalization of a specific system with a specific value of fourfold anisotropy it is necessary to follow the flow line with the specific value of $D$ for this system.\cite{Berlinsky}  This value of $D$ is determined by the initial conditions $x(0),\: Y(0),\text{ and }Y_4(0)$ for the system, where ``(0)" indicates that renormalization begins at the microscopic level where $L=1, \: \ell=\ln L=0$.  To accomplish this, all variables are expressed in terms of the initial value $x(0)$.  In the case of $Y(0) \text{ and }Y_4(0)$, this requires using the definition of $x$ to substitute for the temperature:
\begin{equation}
k_BT=\frac{\pi J}{x(0)+2}.
\end{equation}
In addition, the number of free parameters is reduced by introducing the ratio of the microscopic fourfold anisotropy to the bare exchange, $\lambda = h_4 /J$, so that 
\begin{equation}
Y_4(0) = \lambda \: \frac{(x(0)+2)}{2\pi}\: e^{\frac{-2\pi}{x(0)+2}}.
\label{Y4(0)}
\end{equation}
Since $y_0$ is a phenomenological parameter, its value is not known at the microscopic level.  To find a reasonable estimate for $Y(0)$, the fugacity is expressed instead in terms of the energy of a vortex core, $E_c$. 
\begin{equation}
Y(0) \approx e^{-\frac{E_c}{k_B T}}.
\end{equation}
 An approximate expression for the energy of a vortex core, derived from the Villain model\cite{Berlinsky} is $E_c = \pi J\ln{(\gamma \sqrt{\pi})}\approx 3.6J$. (Here only, $\gamma$ is Euler's constant.)  Then, for the purpose of tuning the system,
 \begin{equation}
Y(0) \approx  e^{-\frac{3.6}{\pi}(x(0)+2)}.
\label{Y(0)}
\end{equation}

When the initial conditions fall on the separatrix ($t=0$) for a system with a particular value of the microscopic anisotropy, the flow proceeds to the fixed point and the corresponding value of $D$ can be identified.  According to eq.(\ref{sep}), this occurs when
\begin{equation}
x(0) =\pm \gamma [ Y(0) - Y_4(0)].
\label{x(0)}
\end{equation}
After substituting from eq.(\ref{Y(0)}) and (\ref{Y4(0)}), this equation can be  solved for $x_\lambda (0)$ for the value of $\lambda$ that characterizes the bare system, and the corresponding value of  $D_\lambda$ can be found using eq.(\ref{Ddef}):
\begin{equation}
D_\lambda = Y_\lambda (0)\,Y_{4,\lambda}(0) .
\label{Dlambda}
\end{equation}

The upper and lower root in eq.(\ref{x(0)}) correspond to whether or not $Y_\lambda(0) > Y_{4,\lambda}(0)$.  The condition $Y_\lambda(0)=Y_{4,\lambda}(0)$ divides these cases, and represents a system with initial conditions at the fixed point at $x=0$. It presumably stays at the fixed point under geometric scaling.

Initial conditions for systems with a wide range of $\lambda$ are collected in Table \ref{Dvalues}.  It turns out that only the positive root of eq.(\ref{x(0)}) is relevant, as this root produces $x_\lambda (0)$ up to $\lambda =7.3$.  Under the assumptions of the generalized Villain model\cite{Jose2}, the maximum value of $y_4$ is unity.  According to eq.(\ref{Ysdef}), this implies a maximum value of $\sqrt{D_\lambda}=\exp{(-\pi)},\: D \approx 1.9\times 10^{-3}$.  Therefore, table entries for $\lambda > 1.0$ are certainly not well-founded.  Ultrathin metallic films on single-crystal substrates have anisotropies with the order of magnitude $10^{-2}>\lambda >10^{-3}$, giving $2\times 10^{-6} < D < 2\times 10^{-5}$.  This is well within the range of validity of the generalized Villain model.
\begin{table}
\caption{\label{Dvalues}For a given ratio $\lambda$ of the microscopic fourfold anisotropy and bare exchange, the value of the initial conditions $x_\lambda (0), Y_\lambda (0)$ and $Y_{4,\lambda}(0)$ given by eq.(\ref{x(0)}), (\ref{Y(0)}) and (\ref{Y4(0)}), such that the system begins on the separatrix are tabulated.  From this initital condition, the system follows the scaling flow line defined for the value $D=D_\lambda$ given by eq.(\ref{Dlambda}). }
\begin{ruledtabular}
\begin{tabular}{|l|l|l|l|l|}
$\lambda= h_4/J$     &      $x_\lambda(0)$        &  $Y_\lambda(0)$ &   $Y_{4,\lambda}(0)$       &     $D_\lambda$     \\ \hline
7.3     &    0   &   0.101 &   0.101   &1.02$\times10^{-2}$   \\ \hline
3.0     &    0.215 & 0.0790 &   0.0620    & 4.90$\times10^{-3}$  \\ \hline
1.0    &     0.421  &   0.0625   &   0.0287   &  1.79$\times10^{-3}$  \\ \hline
0.5    &    0.506 &   0.0566  &  0.0163     &  9.20$\times 10^{-4}$  \\ \hline
0.1    &   0.595 &   0.0512   &   0.00366 &1.87$\times10^{-4}$  \\ \hline
0.01  &   0.620  &  0.0497     & 3.81$\times 10^{-4}$  &1.89$\times10^{-5}$  \\ \hline
0.001 &   0.623  &  0.0495    &3.81$\times 10^{-5}$  &1.89$\times10^{-6}$  \\ \hline
0 & 0.623 &  0.0495  & 0 & 0 \\
\end{tabular}
\end{ruledtabular}
\end{table}

\subsection{Finite-size transition}
To find the coupling for an anisotropic finite system of size $L$, eq.(\ref{xofl}) must be integrated up to $\ell = \ln{L}$.  This is accomplished by solving the quadratic equations in eq.(\ref{sep}) for $Y(\ell)$ and $Y_4(\ell)$, and substituting back into eq.(\ref{xofl}) to give the integral equation
\begin{equation}
-\int d\ell = \int_{x_i}^{x_f} \frac{dx}{\sqrt{x^2+\alpha t}\;\sqrt{x^2+\alpha t+4\gamma^2 D}}.
\label{finite}
\end{equation}
A standard transformation shows that this is an elliptic integral of the first kind\cite{handbook}.  It is solved exactly in terms of the Jacobi elliptic functions in appendix A.

For an approximate solution for small values of $D$, such as those appropriate for ultrathin films, it is useful to rewrite eq.(\ref{finite}) as the difference of squares.
\begin{equation}
-\int d\ell = \int_{x_i}^{x_f} \frac{dx}{\sqrt{(x^2+\alpha t+2\gamma^2 D)^2-(2\gamma^2 D)^2}}.
\label{approx}
\end{equation}
For small enough $D$, an excellent approximation\footnote{The validity of this approximation is clear \textit{a posteriori} from the fact that  in a finite-size transition both $x$ and $\alpha t$ are not simultaneously small in comparison to $D$; that is, the system does not get close to the critical point. } is to neglect the contribution from the term $(2\gamma^2 D)^2$. (See section III. D.) Then, using the substitution $\nu = \sqrt{\alpha t +2\gamma^2 D}/x$ results in an exact differential of arctangent.  As the geometric scaling described by the RG equations removes the sensitivity to the initial value of $x$, let $x_i \rightarrow +\infty$.  Then
\begin{equation}
x_f=x(\ln{L})= \frac{\sqrt{\alpha t + 2\gamma^2 D}}{\tan{[\ln L\ \sqrt{\alpha t + 2\gamma^2 D}]}}.
\label{xapprox}
\end{equation}
This functional form is identical to that found by Bramwell and Holdsworth\cite{Bramwell1} for the isotropic system, with the substitution
\begin{equation}
\alpha t \rightarrow \alpha t + 2\gamma^2 D \equiv \omega,
\label{wdef}
\end{equation}
so that, compared to the isotropic system, the effect of the anisotropy is to shift reduced temperatures by $-2\gamma^2 D$.  

Whereas the second order transition in the infinite system with fourfold anisotropy occurs at the fixed point\cite{Jose2}  $x=0,\: Y=\pm Y_4=\sqrt{D}$, the renormalization flow in the finite system avoids the fixed point so that the condition $x=0$ occurs at the reduced temperature $\omega_{c,0}$.  (The subscript ``c" refers to the critical approximation.) This condition no longer represents a sharp transition, but rather marks when the formation of vortex-antivortex pairs starts to be significant. According to eq.(\ref{xapprox}), $x\rightarrow 0$ when

\begin{equation}
\omega_{c,0} = \alpha t_0 + 2\gamma^2 D= [\frac{\pi}{2\ln L}]^2 
\label{t0}
\end{equation}

Substituting the expression for $x(\ln{L})$ from eq.(\ref{xapprox}) into eq.(\ref{sep}) leads to quadratic equations for $Y(\ln{L})$ and $Y_4(\ln{L})$.  The solutions are most usefully expressed in terms of the scaled variable $\omega/\omega_{c,0}$:
\begin{equation}
\label{xcrit}
x=\frac{1}{\ln{L}} \;\frac{\frac{\pi}{2}\sqrt{\frac{\omega}{\omega_{c,0}}}}{\tan{[\frac{\pi}{2}\sqrt{\frac{\omega}{\omega_{c,0}}}]}}.
\end{equation}
\begin{equation}
P(\omega/\omega_{c,0})=\frac{1}{(\gamma \; \ln{L})^2} \; \frac{(\frac{\pi}{2})^2\frac{\omega}{\omega_{c,0}}}{ \sin^2{[\frac{\pi}{2} \;\sqrt{\frac{\omega}{\omega_{c,0}}}\;]}}.
\end{equation}
Then
\begin{equation}
Y^2 = \frac{1}{2}P(\omega/\omega_{c,0}) + \frac{1}{2}\sqrt{P^2(\omega/\omega_{c,0})-4D^2},
\label{Yapprox}
\end{equation}
\begin{equation}
Y_4^2 = \frac{1}{2}P(\omega/\omega_{c,0}) - \frac{1}{2}\sqrt{P^2(\omega/\omega_{c,0})-4D^2}.
\label{Y4approx}
\end{equation}

\begin{figure}
\scalebox{.50}{\includegraphics{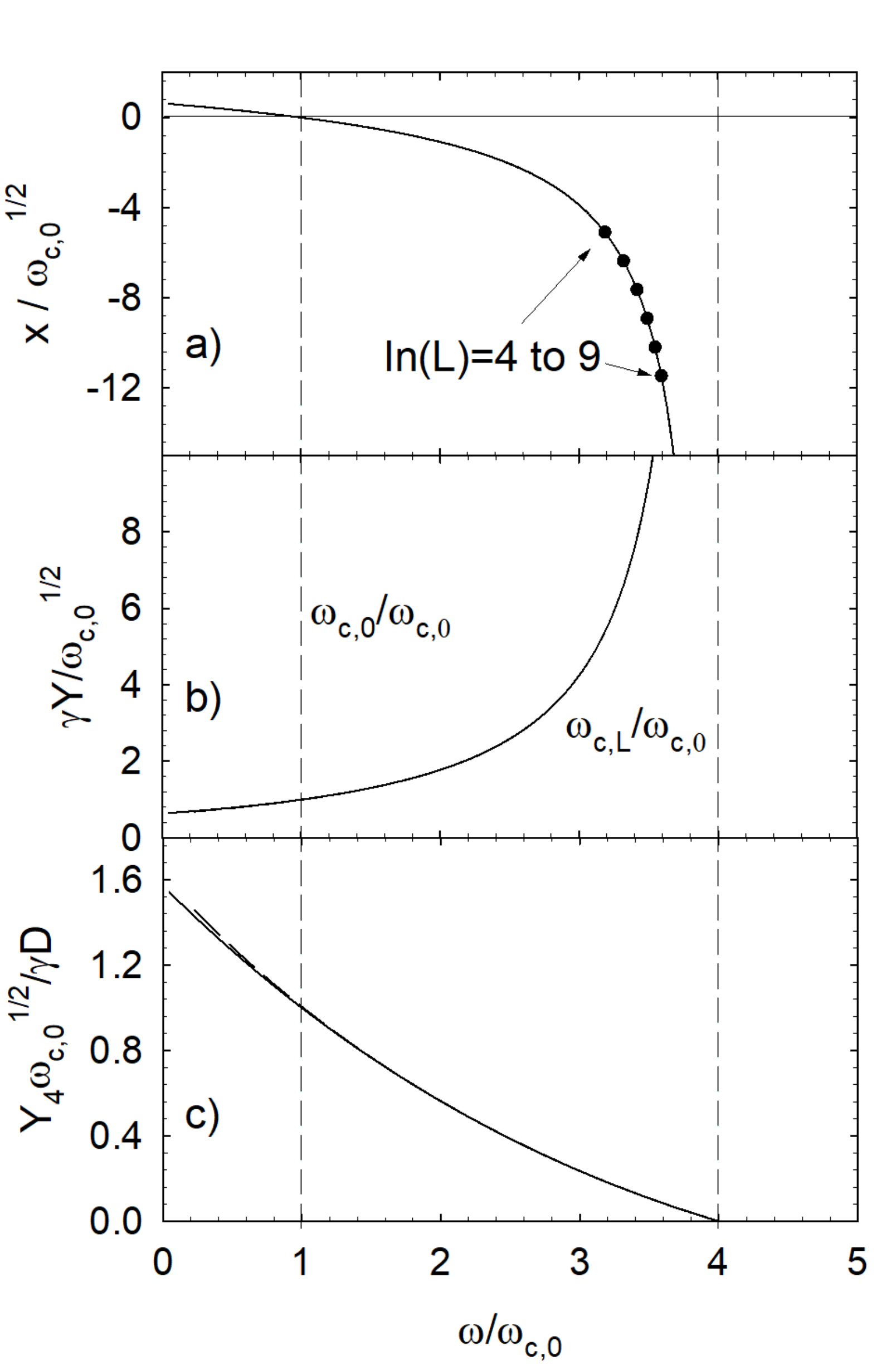}}
\caption{\label{Y4plot} The solutions  in eq.(\ref{xcrit}) to (\ref{Y4approx}) to the RG equations for small anisotropy in the critical approximation are plotted against the scaled variable $\omega/\omega_{c,0}$.  Each variable is normalized so that they they lie on near-universal curves regardless of the values of $\ln{L}$ for the anisotropy parameter $D\leq 1.9\times 10^{-5}$ appropriate for thin ferromagnetic films.  The dashed lines mark $\omega_{c,0}$ where the formation of vortex-antivortex pairs begins, and $\omega_{c,L}$ where the pairs unbind according to the critical approximation.  The plotted variables are: a) the departure of the effective coupling from the critical value, $x=\pi K-2$, b) the fugacity, or vortex density, $Y$, and c) the fourfold anisotropy parameter $Y_4$. The circular dots on the curve in part a) indicate the point beyond which the effective coupling becomes antiferromagnetic for systems of different sizes.}
\end{figure}

These expressions for $x, Y$ and $Y_4$ are plotted in fig.(\ref{Y4plot}), using scaled variables that exhibit (near) universal curves.  Each of the plots has a curve for $\ln{L}=4$ and for $\ln{L}=9$.  Only in part (c) for $Y_4$ is there an indication that the curves do not overlap precisely.  For $D \le 1.9 \times 10^{-5}$, curves generated using the exact solutions in appendix A are indistinguishable within the linewidth from the the approximate solution shown in fig.(\ref{Y4plot}). 

The finite-size transition ends when the system moves to an endpoint where the vortex density gets very large due to the unbinding of the vortex-antivortex pairs, and the anisotropy is screened away.  According to fig.(\ref{Y4plot}), this occurs when $\omega/\omega_{c,0}=4$, independent of system size.  Following Bramwell and Holdsworth\cite{Bramwell1}, this corresponds to $x\rightarrow -\infty$,  and according to eq.(\ref{xapprox}), occurs at\footnote{The notation $t_L$ is used here, rather than $t_C$ as used in ref. 18, to reinforce that the separation of $t_0$ and $t_L$ is a finite-size effect.}
\begin{equation}
\omega_{c,L}=\alpha t_L+2\gamma^2 D = [\frac{\pi}{\ln L}]^2 .
\label{tL}
\end{equation}
In a finite system, the correlation length $\xi$ is limited by the system size.  This implies that the correlation length is maximum near $\omega_{c,L}$, where the paramagnetic vortex gas forms.  In the presence of fourfold anisotropy, eq.(\ref{tL}) can be used to show that it scales as
\begin{equation}
\label{xi}
\xi \sim L=\exp{[\frac{\pi}{\sqrt{\omega}}]} \text{    for   } \omega> \omega_{c,L}.
\end{equation}
Since $D$ is small for thin ferromagnetic films, the difference between this relation and the form $\exp{[\pi/\sqrt{\alpha t}]}$ found for the isotropic system, will be very difficult to observe.

These results confirm that the finite-size KT transition in ultrathin ferromagnetic films survives the inclusion of fourfold anisotropy, and that a second order transition is not expected.  The effective coupling, correlation length, and transition points are those found previously for the isotropic system, if $\alpha t \rightarrow \alpha t + 2\gamma^2 D \equiv \omega$. The principle new finding is the expression in eq.(\ref{Y4approx}) for $Y_4$.

However, it is also clear from fig.(\ref{Y4plot}) that there are important quantitative problems with the calculated system properties.  The effective exchange coupling goes to zero ($x=-2$) before the vortex gas forms at $\omega_{c,L}$ ($x \rightarrow -\infty$).  The condition $K=0$ is marked in fig.(\ref{Y4plot}a) by the circular dots for systems with sizes increasing by integer values of $\ln{L}$ from 4 to 9.  At larger values of $\omega < \omega_{c,L}$ the coupling becomes large and antiferromagnetic, a situation that is unphysical.   Thus the calculation is certainly not reliable near $\omega_{c,L}$, and is unlikely to be reliable outside the region near $\omega_{c,0}$ where $x$ is indeed a small expansion parameter.  Another example  can be seen in part (c) of the figure, where the anisotropy goes to zero with a discontinuity in slope and becomes complex above $\omega_{c,L}$, rather than approaching zero as a smooth and continuous real function.   These are indications that the critical approximation will not be sufficient for a quantitative description of the system across the full temperature range of the finite-size transition, including for the calculation of the magnetic susceptibility.

\subsection{Limiting behaviours in the critical approximation}
Although the critical approximation does not provide a quantitative description throughout the finite-size KT transition, it can provide guidance as to whether or not the system will flow away from the critical point towards a free vortex gas, or towards the critical point and a second order transition\cite{Jose2} that is characterized by either Ising exponents, or non-universal exponents.  This question can be explored using the exact solution in the critical approximation developed in appendix A, where it is shown that the solution in terms of trigonometric functions in eq.(\ref{xapprox}) is replaced by
\begin{equation}
x(\ln{L}) = \sqrt{\alpha t +4\gamma^2 D} \;\: \frac{\text{cn}(B/k , k )}{\text{sn}(B/k,k)}.
\label{exactx}
\end{equation}
sn$(u,k)$ and cn$(u,k)$ are Jacobi elliptic functions\cite{handbook}, and the function parameters for the finite 2DXY model with fourfold anisotropy are
\begin{equation}
\begin{split}
&u= \sqrt{\alpha t +4\gamma^2 D}\: \ln L = \frac{B}{k}, \text{ where}\\
&k= \sqrt{\frac{4\gamma^2 D}{\alpha t +4\gamma^2 D}}\;\; \text{and } B \equiv \sqrt{4\gamma^2 D}\: \ln L.
\end{split}
\label{defB}
\end{equation}

The product of the system size and anisotropy in $B$ characterizes the system behaviour.   The finite-size transition begins at $\alpha t_0$, and the type of phase transition the system undergoes is determined by how it approaches the fixed point ($x=0, \alpha t_0 \to 0$) as a function of the system size. As is outlined in appendix A, $\alpha t_0$ is determined by the condition
\begin{equation}
u = B/k_0 = \kappa(k_0).
\label{k0def}
\end{equation}
In this transcendental equation, $\kappa(k_0)$ is the \textit{complete} elliptic integral of the 1st kind\cite{handbook}, and $k_0 \equiv k(\alpha t =\alpha t_0)$.   These definitions imply that 
\begin{equation}
\alpha t_0 = (1-k_0^2)[\frac{\kappa(k_0)}{\ln{L}}]^2.
\label{at0}
\end{equation}

To understand the critical behaviour, note that the complete elliptic integral can be represented to a high degree of accuracy by the function\cite{handbook}
\begin{equation}
\kappa(k)=m(k)- n(k)\ln{(1-k^2)},
\label{kappapprox}
\end{equation}
where $m(k)$ and $n(k)$ are slowly-varying polynomials with positive values.  Substituting this in eq.(\ref{at0}) and solving for $L$,
\begin{equation}
L = \exp{[\frac{m(k_0)}{\sqrt{\alpha t_0 + 4\gamma^2 D}}]}\: [1-k^2_0]^{[-\frac{k_0 n(k_0)}{2\gamma \sqrt{D}}]}.
\label{product}
\end{equation}
When $k_0$ is small, $\alpha t_0 \gg 4\gamma^2 D$, and the system is in the limit of  small anisotropy.  Then $m(k_0) \approx m(0)=\pi /2$ and the eq.(\ref{product}) becomes
\begin{equation}
L= \exp{[\frac{\pi}{2\sqrt{\alpha t_0 + 4\gamma^2 D}}]}.
\end{equation}
This has the form of eq.(\ref{t0}) and describes a system that approaches the fixed point logarithmically in $L$ and is described by a finite-size KT transition.

When $k_0$ approaches unity, $\alpha t_0 \ll 4\gamma^2 D$ and the system is in the limit of large anisotropy. Then $n(k_0)\approx n(1)=1/2$ and the singular part of eq.(\ref{product}) is
\begin{equation}
\xi \sim L= [\frac{\alpha t_0}{4\gamma^2 D}]^{-\frac{1}{4\gamma \sqrt{D}}}.
\end{equation}
This power law behaviour indicates a second order transition with critical exponent $\nu$.  To find $\nu$, recall that $\sqrt{D}$ is the value of $Y_4$ at the fixed point where $k_0 =1$.  In the limit of large anisotropy in the generalized Villain model, $y_4=1$, so that
\begin{equation}
\nu= \frac{1}{4\gamma\sqrt{D}}=\frac{e^\pi}{4\gamma}= 0.46
\label{nu}
\end{equation}
This value agrees with the analysis of Taroni \textit{et al.}\cite{Taroni} for an infinite system, and represents the 2D Ising limit of the generalized Villain model .  The fact the the correct 2D Ising value $\nu=1$ is underestimated is due to the limitation of the Villain model to small anisotropy\cite{Jose2}, and does not affect the conclusion that this is the Ising-like limit of the model.
 \begin{figure}
\scalebox{.55}{\includegraphics{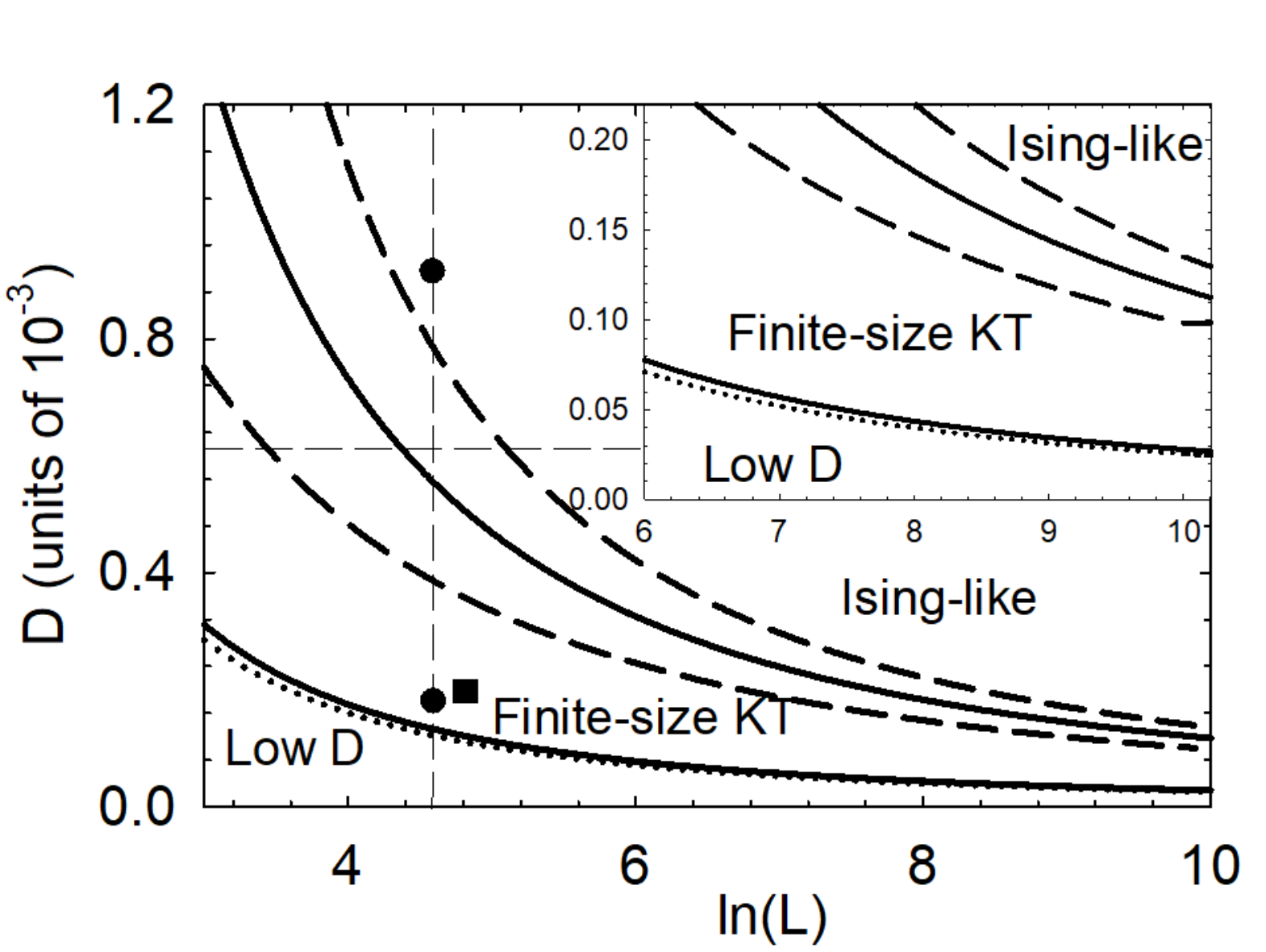}}
\caption{\label{limits} An approximate phase diagram for the 2DXY model wth fourfold anisotropy is plotted in the phase space of the system anisotropy (represented by the paramter D) and system size (represented by the parameter $\ln{L}$).  The solid line near the centre of the figure divides the upper right region where an Ising-like transition occurs, and the lower left region where a finite-size KT transition occurs.  The dividing line is bracketed by two dashed lines that indicate an approximate region where the system crosses over from one behaviour to the other.  In the extreme lower left corner, another solid line marks off a region  ``Low D" where finite-size KT transitions can be treated in the approximation given in eq.(\ref{approx}). The dotted line bounds a region that is essentially the same as the Low D region, where the perturbative approximation discussed in the next section is valid.  The symbols and dashed straight lines are discussed in the text. }
\end{figure}

The boundary where the system crosses over from 2nd order to finite-size KT behaviour depends upon what deviation from ``pure" limiting behaviour is acceptable before the system is considered to be in a crossover region.  A pragmatic approach is to begin with the condition where each of the product terms in eq.(\ref{product}) are equal, or equivalently, when the two terms in eq.(\ref{kappapprox}) are equal.  This occurs when\cite{handbook} $k_0 = 0.967$.   Eq.(\ref{k0def}) then indicates $B=2.72$.  Because of the definition of $B$, this condition defines a line in the parameter space of $D$ vs. $\ln{L}$, and is marked by a solid line near the centre of fig.(\ref{limits}), dividing regions labelled ``Ising-like" and ``Finite-size KT".  

To investigate the power law behaviour moving toward the the boundary line from the region of large $D$ and $\ln{L}$, recognize that $1-k^2_0$ is a small parameter, and set
\begin{equation}
k_0=\sqrt{1-(1-k^2_0)}\approx 1-\frac{1}{2}(1-k^2_0)
\end{equation}
Substituting this and the association developed in eq.(\ref{nu}) into the divergent term in eq.(\ref{product}) gives
\begin{equation}
\begin{split}
\xi &\sim L = (1-k^2_0)^{-\nu}\; (1-k^2_0)^{\frac{\nu}{2}(1-k^2_0)}\\
 & \approx (\frac{\alpha t_0}{4\gamma^2 D})^{-\nu}\;[1+\frac{\nu}{2}(\frac{\alpha t_0}{4\gamma^2 D})\ln{(\frac{\alpha t_0}{4\gamma^2 D})} + ...\;].
\end {split}
\label{tlnt}
\end{equation}
The correction term in eq.(\ref{tlnt}) is of the form $\alpha t_0 \ln{\alpha t_0}$. It increases in size moving toward the boundary from large $D$ and $\ln{L}$, and makes a correction of about -9\% at the boundary when a conservative value of $\nu =1$ is used.  This represents a significant departure from pure power law behaviour, and is expected to cause an effective critical exponent $\nu_{eff} > \nu$ to better describe the system.    The size of the change in $\nu_{eff}$ is difficult to determine in the present analysis, but it is consistent with the boundary lying within the crossover region. A dashed line further into the region labelled ``Ising-like" is included in fig.(\ref{limits}) to indicate an approximate upper bound to the crossover region.  It is set somewhat arbitrarily by the condition that the power law term in eq.(\ref{product}) is twice the size of the exponential term.

The finite-size KT behaviour at the boundary line is investigated by assessing deviations from the correlation length given in eq.(\ref{xi}).  From appendix A, $B/k_L = 2\kappa (k_L)$, so that
\begin{equation}
\alpha t_L = (1-k_L^2)[\frac{2\kappa(k_L)}{\ln{L}}]^2.
\label{atL}
\end{equation}
$\kappa(k_L)$ is expanded\cite{handbook} about $k_L=0$ in powers of $k_L^2$, and terms up to $k^2_L$ on the right side of eq.(\ref{atL}) are retained.  The resulting quadratic equation for $\alpha t_L$ is solved and the term in $\ln{L}$ is isolated.  This yields  
\begin{equation}
\begin{split}
 L= &\exp{[\frac{\pi}{\sqrt{\omega_L}}\; \sqrt{\frac{\omega^2_L}{\omega^2_L -(2\gamma^2 D)^2}}\;]}\\
\xi \sim &L  =  \exp{[\frac{\pi}{\sqrt{\omega}}\; \frac{(1-\frac{1}{2}k^2)}{\sqrt{1-k^2}}]} \: \: ;k<k_L.
\end{split}
\label{xiL}
\end{equation}
For the boundary at $B=2.72$ in fig.(\ref{limits}), $k_L=0.712$ and the characteristic form of the correction length at a finite-size KT transition is modified by an increasing amount as the boundary is approached from the region of low $D$ and $\ln{L}$.  The correction is +7\% at the boundary.  This is of a similar magnitude to the modification of the Ising-like correlation length at the boundary, and suggests that the boundary in fig.(\ref{limits}) is indeed roughly in the centre of the crossover region.   A dashed line further into the region labelled ``Finite-size KT" is included in fig.(\ref{limits}) to indicate an approximate lower bound to the crossover region.  Again, it is set by the condition that the exponential law term in eq.(\ref{product}) is twice the size of the power law term.

In the infinite system, there is no KT behaviour, but rather the 2nd order transition persists to low anisotropy\cite{Jose2}, where it exhibits non-universal critical exponents with $\nu \sim 1/h_4$.  The exact solution of the RG equations in the critical approximation indicates that this type of transition no longer exists in the finite system, except perhaps as a qualitative explanation of the continuously varying effective critical exponents in the transition region.
\footnote{
As $k_0 \to 1$, eq.(\ref{exactx}) can be expanded in hyperbolic functions as
\begin{equation*}
\lim_{x \to 0}x = \frac{\sqrt{\alpha t_0 + 4\gamma^2 D}}{\sinh u}\times [1-\frac{1-k_0^2}{4}\cosh^2{u}+ ...\;].
\label{expand}
\end{equation*}
The second root leads to the Ising-like transition, and this root exists even when $\ln{L}$ is finite.  The first root $\sinh{u} \to \infty $ exits only when $\ln{L} \to \infty$, and can be shown to be the root for a 2nd order transition at low anisotropy.  In addition, the product of the roots is dominated by the stronger, Ising-like root, so that the transition at low anisotropy is not expressed in a finite system.
}

A separate, but closely related, question is defining the parameter space where the description of the $\textit{entire}$ finite-size KT transition in the critical approximation is consistent with the results of the previous section.  That is, what is the  ``Low D" region within which the approximation leading from eq.(\ref{approx}) to (\ref{xapprox}) is valid at $k_0$ rather than at $k_L$?  Since eq.(\ref{atL}) and (\ref{at0}) differ only in having $\pi \to \pi /2$ and subscripts $L \to 0$, the analysis in eq.(\ref{xiL}) can be applied directly.  To maintain a similar correction of about 7\% to the calculation of properties for the full transition requires $k_0 = 0.712$.  Using eq.(\ref{k0def}), this gives $B=1.35$ as the boundary.  This second solid line is marked on fig.(\ref{limits}) to define a region in the lower left corner where the Low D method of solving the RG equations is essentially equivalent to the exact solution.  This region certainly contains the parameters describing ultrathin ferromagnetic films.

The dotted line in fig.(\ref{limits}) is discussed in the next section.

Fig.(\ref{limits}) summarizes a number of important conclusions concerning the observation of a finite-size KT transition in a real, finite system with fourfold anisotropy.  First, the type of transition depends upon the product of the anisotropy and the system size in the combination $B= 2\gamma \sqrt{D} \ln{L}$, not just on the size of the anisotropy alone.  Second, the second order transition with non-universal exponents $\nu \sim 1/h_4$, predicted for the infinite system, does not occur in the finite system.   These observations do not contradict the findings of Taroni \textit{et al.}\cite{Taroni}, who found non-universal effective exponents for low anisotropy in simulations performed using a Monte Carlo technique. These effective exponents arise in the transition region in fig.(\ref{limits}) They used planar spins on a 2D square lattice with $10^4$ sites, so that $\ln{L} = 4.6$. This system size is marked by the vertical dashed line in the figure.  The lower dot on this line marks the value of $D$ for their calculation when $\lambda =0.1$.  This was the largest anisotropy for which the calculated magnetization exhibited finite-size KT behaviour.  The upper dot marks the value of $D$ when $\lambda = 0.5$, the smallest anisotropy for which the calculated magnetization exhibited Ising-like behaviour.  (No calculations are reported for anisotropy between these values.) As can be seen, the Monte Carlo simulations are in good agreement with the present analysis.  

Comparing the values in Table I to those in fig.(\ref{limits}), it is clear that metallic, ferromagnetic thin film systems will always exhibit a finite-size KT transition.  For other types of systems, neutron scattering experiments have been reported for a few magnetic, layered insulators with weak interlayer coupling, so that they behave as a 2DXY system at low temperature.  For ferromagnetic Rb$_2$CrCl$_4$, it is estimated\cite{Taroni} that $\lambda \approx 0.013$ (so that Table I gives $D \approx 2\times 10^{-4}$),  and that the system size is limited by the interlayer coupling to about $\ln{L} \approx 4.8$.  These co-ordinates are indicated by the square in fig.(\ref{limits}), in a region where finite-size KT behaviour is expected, in agreement with the analysis of the neutron scattering measurements.  A second example is antiferromagnetic K$_2$FeF$_4$, for which anomalous exponents have been observed.\cite{Thurlings}  For this compound, $\lambda$ is estimated to be 0.33 due to gaps in the magnon spectrum\cite{Taroni}.  The corresponding value of $D$ is indicated by the horizontal dashed line in the figure.  The authors of the neutron scattering analysis\cite{Thurlings} argue that the observed critical exponents are comparable to those of the 2D Ising model.  According to fig.(\ref{limits}), this would imply a large system size of $\ln L \sim 6$, despite the interlayer interactions that limit the range of 2D magnetic behaviour.  On the other hand, Taroni \textit{et al.}\cite{Taroni} point out that the measured value of $\beta = 0.15$ is intermediate between the 2D Ising value (0.125) and the effective value for the finite 2DXY model (0.231).  This suggests that  the effective 2D system size is smaller, and that antiferromagnetic K$_2$FeF$_4$ sits within the transition region, consistent with fig.(\ref{limits}).
 
\section{Renormalized coupling across the finite-size transition in the presence of fourfold anisotropy} 
\subsection{Flow equations}

In order to explore the entire temperature range of the finite-size transition, it is better to retain the original RG equations and work directly with a normalized, effective coupling $\pi K/2 \equiv \delta=1+x/2$ within the range $\delta >0$.  Then the RG equations are:
\begin{equation}
\label{RGdelta}
Y=y_0 e^{-\pi \delta} \:\:,  Y_4=y_4 e^{-\frac{\pi}{\delta}}
\end{equation}
\begin{equation}
\frac{d\delta}{d\ell} = \frac{1}{2}\gamma^2\: ( -\delta^2Y^2 + Y_4^2),
\label{dofl}
\end{equation}
\begin{equation}
\frac{dY}{d\ell} = -2(\delta -1) Y
\label{Ydofl}
\end{equation}
\begin{equation}
\frac{dY_4}{d\ell} = \frac{2(\delta -1)}{\delta} Y_4.
\label{Y4dofl}
\end{equation}
Expressed in these variables, the flow equations have a fixed point at $\delta=1, \; Y=\pm Y_4$.

These equations are not amenable to a closed solution.  However, an approximation is suggested by the exact solution in the critical approximation in appendix A.  Because the renormalization flow approaches the fixed point logarithmically in $L$ (see eq.(\ref{xcrit})), finite systems do not get very close to the fixed point.   Rather, they follow a path where $(Y_4/Y)^2 \ll 1$ in the pertinent range of reduced temperature.  From eq.(\ref{Yofk}) and (\ref{Y4ofk}),
\begin{equation}
\frac{Y_4}{Y}=\frac{1-\text{dn}(\frac{B}{k},k)}{1+\text{dn}(\frac{B}{k},k)},
\end{equation}
where dn($u,k$) is the third Jacobi elliptic function\cite{handbook}.  This ratio has its largest value when $t=0$, or $k=1$.  In this limit dn($u,1$) $\rightarrow$ sech$u$, and 
\begin{equation}
B = \text{arccosh}[\frac{1+\sqrt{\epsilon}}{1-\sqrt{\epsilon}}].
\end{equation}
When $\epsilon = (Y_4/Y)^2$ is small, the term in $Y_4^2$ in eq.(\ref{dofl}) can be neglected, and the resulting solutions for $\delta(\ell)$ and $Y(\ell)$ can be used to find $Y_4(\ell)$ as a perturbation through the ratio of eq.(\ref{Ydofl}) and (\ref{Y4dofl}).  Choosing $\epsilon = 0.1$ gives $B$=1.27 as the upper limit.  The resulting boundary for this approximation is shown on fig.(\ref{limits}) as the dotted line, where it almost overlaps the boundary for the ``Low D" solution.  This gives a sizeable parameter space where this approach is valid, and certainly includes the ferromagnetic thin films that are the focus of the present analysis.

This approach will be termed the ``perturbative approximation". The relevant equations in this approximation are
\begin{equation}
\frac{d\delta}{d\ell} = -\frac{1}{2}\gamma^2\: \delta^2Y^2,
\label{daofl}
\end{equation}
\begin{equation}
\frac{dY}{d\ell} = -2(\delta -1) Y
\label{Ydofl2}
\end{equation}
\begin{equation}
\frac{dY_4}{Y_4}=-\frac{dY}{\delta Y}.
\label{Yratio}
\end{equation}
  It is important to reiterate that although eq.(\ref{daofl}) to (\ref{Yratio}) do not display the fixed point of the original RG equations, they are a very good approximation in the region some distance from the fixed point where the renormalization flow carries a finite system. 

Rearranging eq.(\ref{Ydofl2}) as an expression for $Y$, and substituting for one power of $Y$ in eq.(\ref{daofl}), leads to a relation between exact differentials that can be integrated as
\begin{equation}
\ln \delta + \frac{1}{\delta} = \frac{\gamma^2}{8}Y^2 + \frac{C}{8}.
\label{findC}
\end{equation}
The integration constant $C/8$ can be identified by expanding this equation about $\delta=1$ and comparing to eq.(\ref{sep}) in the region where they are both valid. This gives
\begin{equation}
C= 2\gamma^2 D + \alpha t -8 = \omega -8.
\label{defC}
\end{equation}
Incorporating this in eq.(\ref{findC}), provides a final expression for the fugacity in the perturbative approximation:
\begin{equation}
\gamma^2 Y^2 = 8(\ln \delta + \frac{1}{\delta} -1) +\omega.
\label{fugacity}
\end{equation}

\subsection{Finite-size transition}

\begin{figure}
\scalebox{.58}{\includegraphics{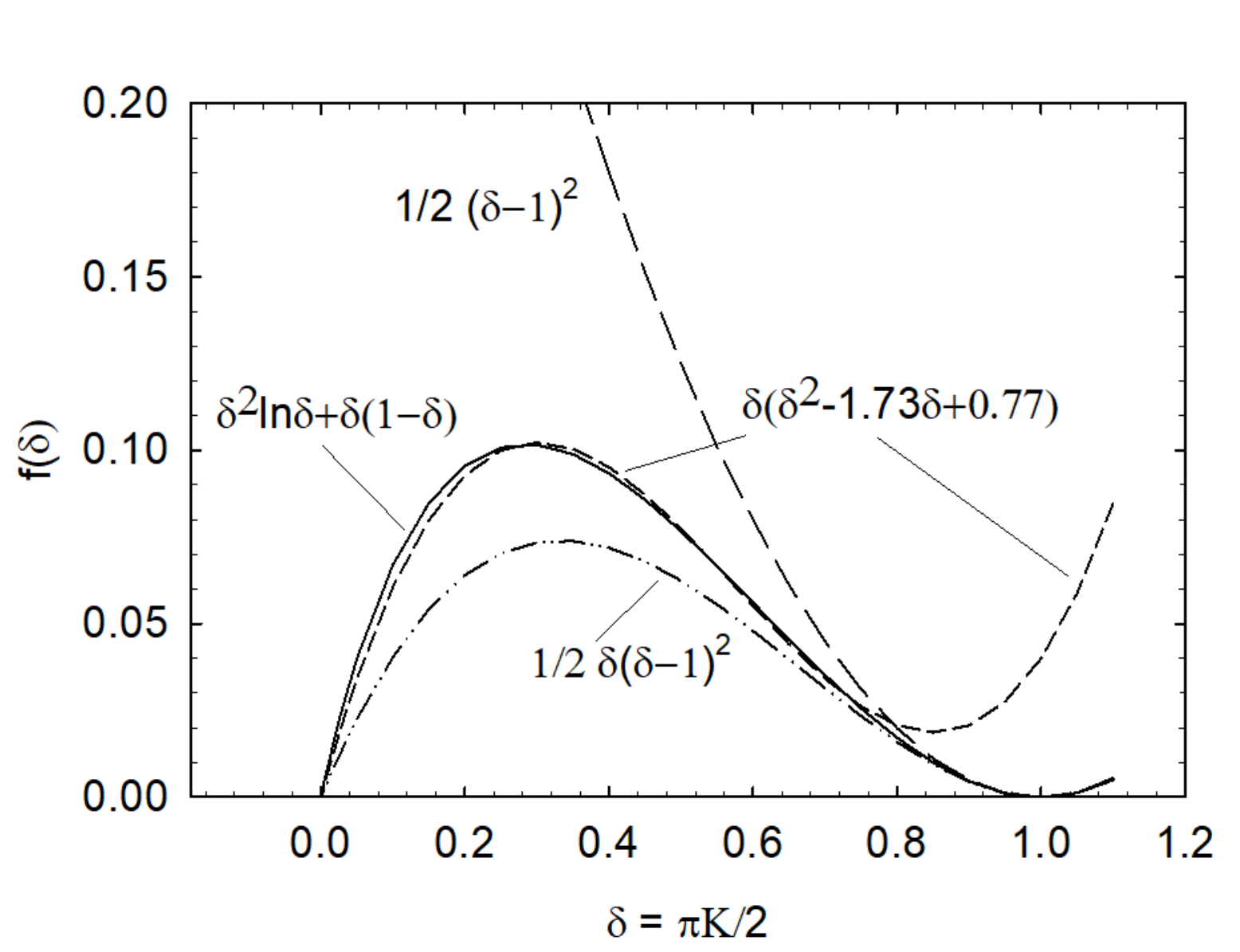}}
\caption{\label{f(delta)}The function $f(\delta )$ in the square brackets of eq.(\ref{dfinite}) is illustrated, along with various approximations to it.  The solid line indicates the exact function.  The dashed line $1/2\;(\delta -1)^2$ is the quadratic function used in the critical approximation in Section III.  The remaining two lines are piecewise approximating polynomials for the regions $\delta > 3/4$ and $0< \delta < 3/4$. }
\end{figure}

Substituting the expression for $Y^2$ from eq.(\ref{fugacity}) into eq.(\ref{daofl}), and separating variables gives
\begin{equation}
\begin{split}
-d\ell &= \frac{d\delta}{4[\delta^2 \ln \delta +\delta (1-\delta) ]+\frac{1}{2}\omega \delta^2} \\
&\equiv \frac{d\delta}{4f(\delta)+\frac{1}{2}\omega \delta^2}
\end{split}
\label{dfinite}
\end{equation}
Because of the presence of $\ln \delta$, this integral cannot be performed analytically.  However, the denominator is well-behaved so long as $\delta >0$, and various approximations are instructive.  These approximations are illustrated in fig.(\ref{f(delta)}), where the term $f(\delta)$ in square brackets in eq.(\ref{dfinite}) is plotted. The solid line is the exact function, and the purely quadratic function $1/2\;(\delta-1)^2$ is the critical approximation for small $x\: (\delta \approx 1)$ used in the previous section.  The figure makes it clear why this approximation is unreliable near $\delta \approx 0$, where the free vortex gas forms.

A simple and effective representation of $f(\delta)$ is achieved by piecewise polynomials.  In the range $\delta \ge 3/4$, the polynomial $1/2\;\delta (\delta -1)^2$ is used.  Then eq.(\ref{dfinite}) has the form
\begin{equation}
\int_{\delta_i}^{\delta_f} \frac{d\delta}{\delta X(\delta)} = - \ln L,
\label{larged1}
\end{equation}
where $X(\delta) = a +b\delta + c\delta^2$ with $a=2,\: b=-4+\omega/2$ and $c=2$. This integral is given in appendix B in eq.(\ref{exactd}).  Its qualitative form is more easily seen in the limit $\omega \ll 4$, where  the expression simplifies to
\begin{equation}
\delta=1+\frac{\sqrt{\omega}}{2\tan{[ \sqrt{\omega}\ln{[L\;(\frac{\delta^2}{(1-\delta)^2+\frac{\omega\delta}{4}})^{\frac{1}{4}}]}]}}.
\label{larged2}
\end{equation}

\begin{figure}
\scalebox{.50}{\includegraphics{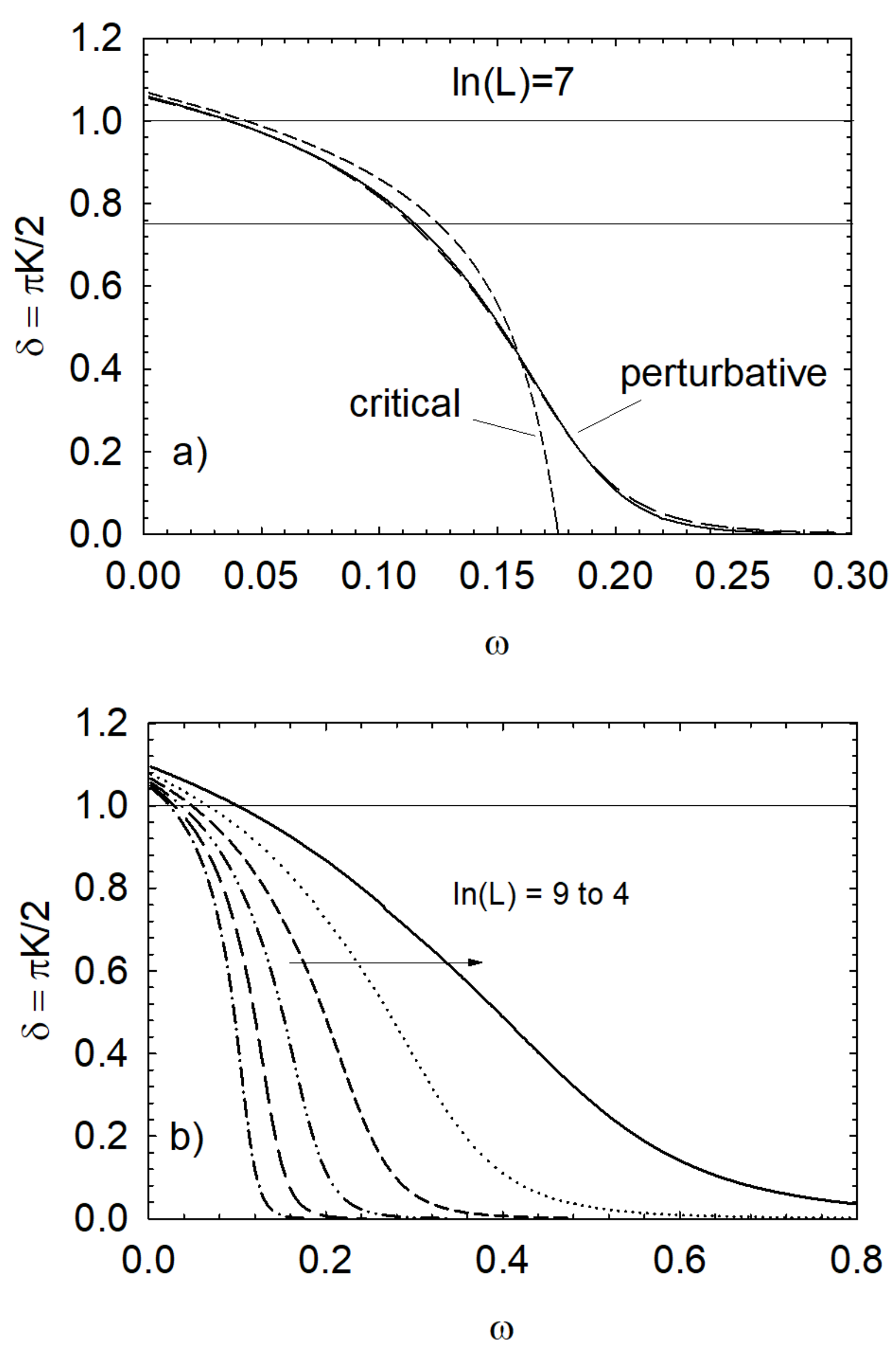}}
\caption{\label{delta}a) The normalized effective coupling $\delta(\omega)$ is plotted for both the critical and perturbative approximations to the RG equations, for the case $\ln{L}=7$ and $D=1.9\times 10^{-5}$.  For the latter, the solid line is a numerical integration of eq.(\ref{dfinite}) and the long dash line that almost overlaps with the solid line is the polynomial approximation that gives eq.(\ref{exactd}) and (\ref{smalld}).  The result in the critical approximation is given by eq.(\ref{xapprox}).  b) The polynomial approximation is used to plot $\delta(\omega)$ for the same value of $D$ and a range of system sizes $L$. }
\end{figure}
 
This coupling is closely related to eq.(\ref{xapprox}) in the critical approximation, but represents an important qualitative change.  While there is only a modest difference near $\delta\approx1$, the behaviour at $\delta\approx 0$ is very different.  The implicit equation for $\delta$ includes a term in $\ln\delta$ that rules out solutions for $\delta \le 0$.  Thus the system approaches the limit of vanishing exchange coupling ($\delta =0$) asymptotically in $\omega$.

In the range $3/4 \ge \delta>0$ in fig.(\ref{f(delta)}), $f(\delta)$ is reasonably approximated by the polynomial $\delta(\delta^2-1.73\delta+0.77)$.  This is again of the form in eq.(\ref{larged1}), but the algebra is more complicated.  The result is given in appendix B in eq.(\ref{smalld}), and is qualitatively similar to eq.(\ref{larged2}) with the substitution of a generalized expression for
\begin{equation}
\ln L \rightarrow \ln L(\omega)=\ln L + g(\omega), 
\end{equation}
where $g(\omega)$ is a function that arises from matching the two polynomial approximations at $\delta=3/4$, and is given in eq.(\ref{L(w)}).
 
These various expressions for $\delta(\ln{L})$ are compared in fig.(\ref{delta}a) as a function of $\omega$ for parameters appropriate for ultrathin Fe/W(001) films\cite{Atchison}$^,$\footnote{The value of $D$ is relevant only in that it is small enough for the approximation in eq.(\ref{daofl}) to (\ref{Yratio}) to hold, since $D$ and $\alpha t$ occur only in the combination $\omega$.}: $D=1.9\times 10^{-5}$ and $\ln L =7$.    The solid curve is a numerical integration of the relation in eq.(\ref{dfinite}).  The polynomial approximation to it is given by the long dash line.  As can be seen, the closed expressions in eq.(\ref{exactd}) and (\ref{smalld}) reproduce the exact result very well. The two curves nearly overlap; the deviation is greatest at the matching point $\delta=3/4$ and for large $\omega$ as $\delta \rightarrow 0$.  The effective exchange coupling approaches $\delta=0$ asymptotically, as is appropriate for a finite system, and there is no region where the coupling becomes antiferromagnetic.  

The result in the critical approximation, given by eq.(\ref{xapprox}), is also shown as a short dash line in fig.(\ref{delta}a).   Bramwell \textit{et al.}\cite{Bramwell1} have shown, using the critical approximation, that the effective critical exponent of the magnetization, given by
\begin{equation}
\label{beff}
\frac{\partial [\ln M(t)]}{\partial (\ln t)}|_{\delta=1} = \frac{\partial [\ln M(\delta(t))]}{\partial \delta (t)}|_{\delta=1} \;\frac{\partial \delta(t)}{\partial (\ln t)}|_{\delta=1},
\end{equation}
has a universal value 0.231..., and that this prediction is well supported by experiment.\cite{Taroni}.  Comparing the two approximations, there is a small shift in the region near $\delta =1$, but the slopes of the curves are very nearly the same.\footnote{Compare eq.(\ref{larged2}) and (\ref{xapprox}).}  For this reason, the value of the effective critical exponent will not be affected.  Fig.(\ref{delta}b) shows $\delta (\omega)$ for a range of system sizes.

\subsection{Width of the finite-size transition}
The onset of the formation of vortex-antivortex pairs continues to occur when the value of the coupling is equal to the value at the fixed point ($\delta=1,\;x=0$).  In the perturbative approximation, an expression for $\omega_0$ can be found from eq.(\ref{exactd}), as it applies for $\delta > 3/4$.  As $\delta \rightarrow 1$, a small angle approximation for the tangent at an angle slightly less than $\pi/2$ yields the solution given in eq.(\ref{exactw0}) in appendix B.  When $\omega_0 /8 \ll 1$, this becomes
\begin{figure}
\scalebox{.60}{\includegraphics{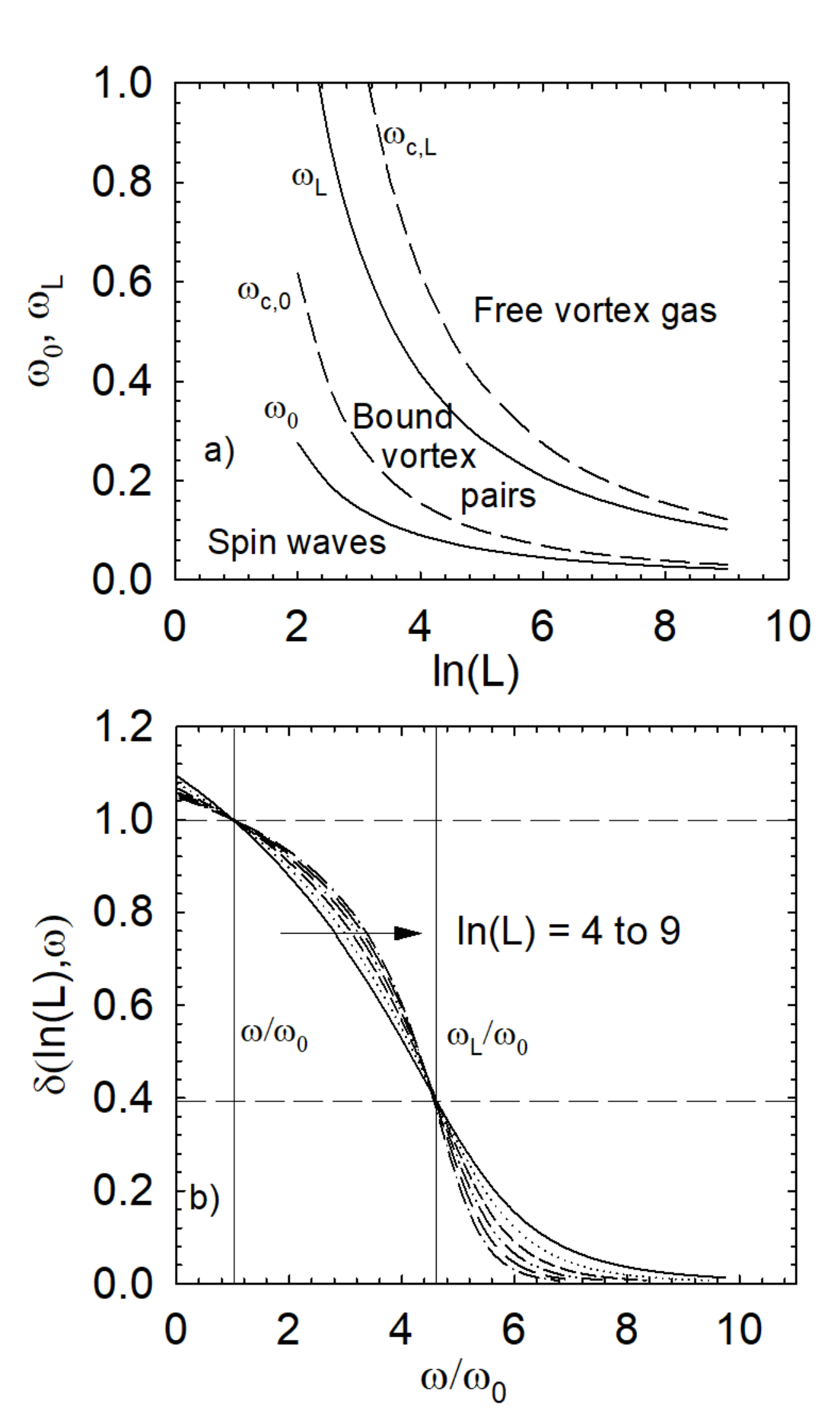}}
\caption{\label{width}a) The finite-size KT transition is spread out over a range of reduced temperature (or $\omega$) that depends upon the size of the system, $L$.  $\omega_0$ divides the condition where spin waves dominate at low temperature and the formation of  bound vortex-antivortex pairs begins.  $\omega_L$ divides the conditions where the vortices and anti-vortices are bound and unbound. The curves defined by the critical and perturbative approximations are shown by dashed and solid lines, respectively.  b) $\delta(\omega)$ for different values of the system size, $L$, are plotted against the normalized parameter $\omega/\omega_0$.  This displays the expected independence upon system size at $\omega_0$, as well as a second scaling point at $\omega_L \approx 4.61\omega_0$ at the inflection points of the curves, where $\delta =0.395$.   }
\end{figure}
\begin{equation}
\frac{\pi}{2\sqrt{\omega_0}} - \frac{1}{4} =  \ln [L\;(\frac{4}{\omega_0})^{1/4}].
\label{lnseries}
\end{equation}

The values of $\omega_0$ are plotted against the system size in fig.(\ref{width}a), using a solid line.  The dashed lines are the results of the critical approximation in eq.(\ref{t0}) and (\ref{tL}).  The scaling properties of $\delta$ in the perturbative approximation are revealed by replotting fig.(\ref{delta}b) as a function of the scaled parameter $\omega/\omega_0$ in fig.(\ref{width}b).  A second scaling point occurs at the point of inflection of all of the curves, at\footnote{Here, the symbol $\beta$ does not represent a critical exponent.} $\omega/\omega_0=4.61... \equiv \beta$ , at which point $\delta=0.395...$ independent of the system size.  For the finite system, this point of steepest descent is all that remains of the instantaneous jump in the coupling observed at $T_{KT}$ in the isotropic, infinite system.  It is identified as $\omega_L$ and the values are plotted in fig.(\ref{width}a) using a solid line.  

The scaling $\omega_L = \beta \omega_0$ can be used with eq.(\ref{lnseries}) to determine the scaling of the correlation length   \begin{equation}
\xi \sim L = (\frac{\omega}{4\beta e})^\frac{1}{4}\; \exp{[\frac{\pi\sqrt{\beta}}{2\sqrt{\omega}}]},\:\: \omega \ge \omega_L.
\label{correl}
\end{equation}
This displays the exponential singularity associated with a KT transition.  The constant in the exponential factor has been altered by a factor of $\sqrt{\beta}/2=1.07$ from that found in the critical approximation. The prefactor to the exponential has no singularity and does not affect the scaling behaviour substantially within the range $\omega \ge \omega_L$ where the estimate applies.

\subsection{Screening of the anisotropy}

With solutions for $\delta(\omega)$ and $Y(\omega)$ derived under the condition that $Y_4^2 \ll Y^2$, it is now possible to solve for $Y_4(\omega)$ as a perturbation using eq.(\ref{Yratio}).  First, note from eq.(\ref{fugacity}) that $\gamma Y$ is the square root of the function $h(\delta)$, where
\begin{equation}
h(\delta)=8(\ln\delta +\frac{1}{\delta}-1)+\omega.
\end{equation}
Using this relation, eq.(\ref{Yratio}) can be written as
\begin{equation}
\frac{dY_4}{Y_4}=-\frac{1}{4} \delta(\ell) \frac{dh[\delta(\ell)]}{d\delta(\ell)} d\ell.
\end{equation}
As the variables $Y_4$ and $\ell$ are separated, integration leads to
\begin{equation}
Y_4(\ln L, \omega) = A \exp{[\;-2\int_0^{\ln L}(\frac{1}{\delta(\ell,\omega)}-1)\;d\ell\;]}.
\label{Y4ofw}
\end{equation}

The integration constant $A$ can be determined in the limit $\delta \rightarrow 1$, where the critical and perturbative approximations are both valid.  Then, from eq.(\ref{Ddef}),
\begin{equation}
\begin{split}
Y_4(\ln L, \omega_0)&=\frac{D}{Y(\ln L, \omega_0)}= \\
&=A \exp{[\;-2\int_0^{\ln L}(\frac{1}{\delta(\ell,\omega_0)}-1)\;d\ell\;]},
\end{split}
\end{equation}
since $\delta=1$ at $\omega_0$.  This notation is understood to mean that $\omega_0$ is a constant in the integral and takes the value appropriate for the system size of the endpoint $L$.  According to eq.(\ref{fugacity}),
\begin{equation}
Y(\ln L, \omega_0)=\frac{\sqrt{\omega_0}}{\gamma}.
\end{equation}
Using these results, eq.(\ref{Y4ofw}) can be written as
\begin{equation}
Y_4(\ln L, \omega) = Y_4^0\; \exp{[\;-2\int_0^{\ln L}\frac{d\ell}{\delta(\ell,\omega)}\;]},
\label{Y4exact}
\end{equation}
\begin{equation}
\text{with }  Y_4^0 = \frac{\gamma D}{\sqrt{\omega_0}\;}  \exp{[\;2\int_0^{\ln L}\frac{d\ell}{\delta(\ell,\omega_0)}\;]}.
\label{Y4norm}
\end{equation}

\begin{figure}
\scalebox{.5}{\includegraphics{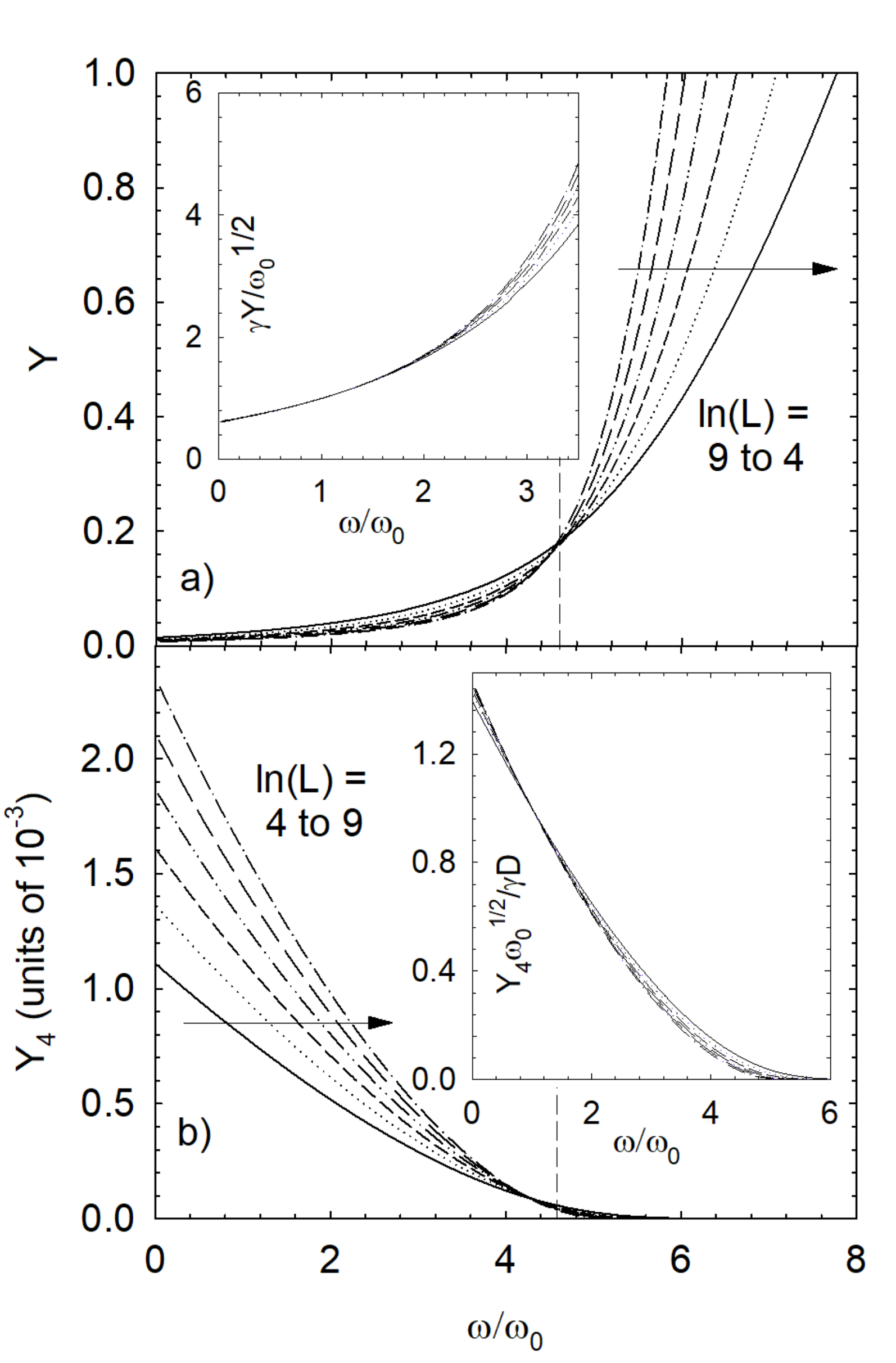}}
\caption{\label{Y(w)}a)  The fugacity, or vortex density, is plotted as a function of $\omega/\omega_0$ using eq.(\ref{fugacity}) and $\delta(\omega)$ from the perturbative approximation. Each individual curve shows no clear marker of the finite-size transition, but $\omega_L$ is indicated by the scaling point and the dashed line.   The inset shows the same curves with $Y$ for different system sizes normalized by $\gamma/\sqrt{\omega_0}$, still plotted against $\omega/\omega_0$.  b) The anisotropy $Y_4$ from eq.(\ref{Y4exact}) for different system sizes is plotted against the scaled parameter $\omega/\omega_0$.  The anisotropy goes to zero smoothly and continuously just beyond $\omega_L$.  In the insert, $Y_4$ is scaled by $\sqrt{\omega_0}/\gamma D$.}
\end{figure}

The fugacity $Y(\omega)$ from eq.(\ref{fugacity}) and the anisotropy $Y_4(\omega)$ from eq.(\ref{Y4exact}) are plotted against $\omega /\omega_0$ in fig.(\ref{Y(w)}a) and (\ref{Y(w)}b), respectively.  Each fugacity curve rises smoothly with no marker of either the beginning ($\omega_0$) or end ($\omega_L$) of the finite-size KT transition,  although the presence of the transition can be seen through the scaling at $\omega_L$. Because the exchange coupling never renormalizes to zero in these finite systems, bound vortex pairs and free vortices always coexist.  The inset shows the same data, again plotted against $\omega/\omega_0$, but now normalized as $\gamma Y/\sqrt{\omega_0}$.  This gives a near universal curve for $\omega/\omega_0 \lesssim 2$, in good agreement with the critical approximation in fig.(\ref{Y4plot}b) in this range. 

The anisotropy $Y_4(\omega)$ plotted in fig.(\ref{Y(w)}b) approaches zero smoothly and continuously, where it has the functional form of an exponential of the exponential function.  In the inset, the anisotropy is scaled by $\sqrt{\omega_0}/\gamma D$ to allow direct comparison with fig.(\ref{Y4plot}c).  The near universal curve for $\omega/\omega_0 \lesssim 2$ observed in the critical approximation is seen here as well.  The anisotropy extends beyond $\omega_L$, especially for smaller system sizes.  This again reflects the fact that smaller systems contain a more truncated distribution of vortex sizes.

\section{The magnetic susceptibility}
\subsection{Contribution due to fluctuations in the scalar magnetization}

Archambault \textit{et al.}\cite{Archambault1} have studied the magnetic susceptibility in a finite-size, isotropic 2DXY system, and demonstrated that a broad peak occurs as the spatial range over which the effective coupling varies diverges until it is limited by the system size.  Their analysis uses the \textit{harmonic} 2DXY model (which is almost equivalent to the Villain model) on a $L\times L$ square lattice of $N$ spins with no explicit fourfold anisotropy term.  The lattice spacing is unity.  They find that the vector magnetization has a well-defined scalar magnitude $M$ in the spin wave region at low temperature.  The magnetization rotates ``slowly" in the isotropic XY plane, so that in finite spin systems the scalar magnetization is a well-defined quantity on experimental time scales despite the lack of anisotropy.  They suggest that when magnetic properties such as the magnetic susceptibility or critical behaviour are measured in an applied field, the field pins the direction of the magnetization so that the relevant fluctuations are in the magnitude of the magnetization.   

Defining the scalar magnetization in terms of the in-plane unit spins $\mathbf{S}_i$ at site \textit{i},
\begin{equation}
M=\frac{1}{N}\sqrt{(\sum_i \mathbf{S}_i )\cdot (\sum_i \mathbf{S}_i )},
\end{equation}
they calculate the susceptibility per spin, $\chi$, as the fluctuations in the scalar magnetization:
\begin{equation}
\chi=\frac{ N}{T} [\langle M^2 \rangle-\langle M\rangle^2].
\end{equation}
They work in units where the Boltzmann constant is unity. In the low temperature spin wave region, the magnetization has the form
\begin{equation}
\langle M\rangle= \exp{(-\frac{G(0)}{2K})},
\end{equation}
where $G(0)$ is the 2D Green's function for the square lattice, evaluated at the origin (see appendix C). The second moment of the scalar magnetization is
\begin{equation}
\langle M^2 \rangle=\frac{1}{N} \sum_r \langle M\rangle^2 \exp{(\frac{G(r)}{K})},
\end{equation}
so that
\begin{equation}
\label{lowT}
\begin{split}
\frac{\chi T}{N} = &  [\frac{1}{N} \sum_r \exp{(-\frac{G(0)}{K})} \exp{(\frac{G(r)}{K})}] \\
& -\exp{(-\frac{G(0)}{K})}.
\end{split}
\end{equation}

This expression is generalized to higher temperature in the range of the finite-size Kosterlitz-Thouless transition by replacing the bare coupling $K$ by the effective coupling $K_{eff}(r,\omega) \equiv 2 \delta(r,\omega)/\pi$ as determined by the renormalization group equations.  A choice must be made for the value $K_{eff}(r)$ to be used in the final term in eq.(\ref{lowT}), as it is outside the sum over $r$.  Because $G(0)$ is dominated by fluctuations at small wavevector, the choice $K_{eff}(L)$ is made. 

Archambault \textit{et al.}\cite{Archambault1} show that a series expansion of the exponential in $G(r)$ converges very quickly.  When only the first term beyond unity is kept, then the susceptibility can be divided into a part $\chi_S$ due to spin waves,
\begin{equation}
\label{chiS}
\frac{\chi_S T}{N} = \frac{1}{N} \sum_r   \frac{\pi^2 G^2(r)}{8\; \delta^2(r)} \exp{(-\frac{\pi G(0)}{2\delta(r)})},
\end{equation}
and a part $\chi_V$ due to vortices,
\begin{equation}
\label{chiV}
\begin{split}
\frac{\chi_V T}{N} =&  \frac{1}{N} [\sum_r \exp{(-\frac{\pi G(0)}{2\delta(r)})}] \\
& -\exp{(-\frac{\pi G(0)}{2\delta(L)})}.
\end{split}
\end{equation}
Because a characteristic experimental thin film system size\cite{Atchison} is $L\approx e^7$, eq.(\ref{chiS}) and (\ref{chiV}) for the susceptibility can be evaluated in the continuum limit.  This is outlined in appendix C.  
\begin{figure}
\scalebox{0.6}{\includegraphics{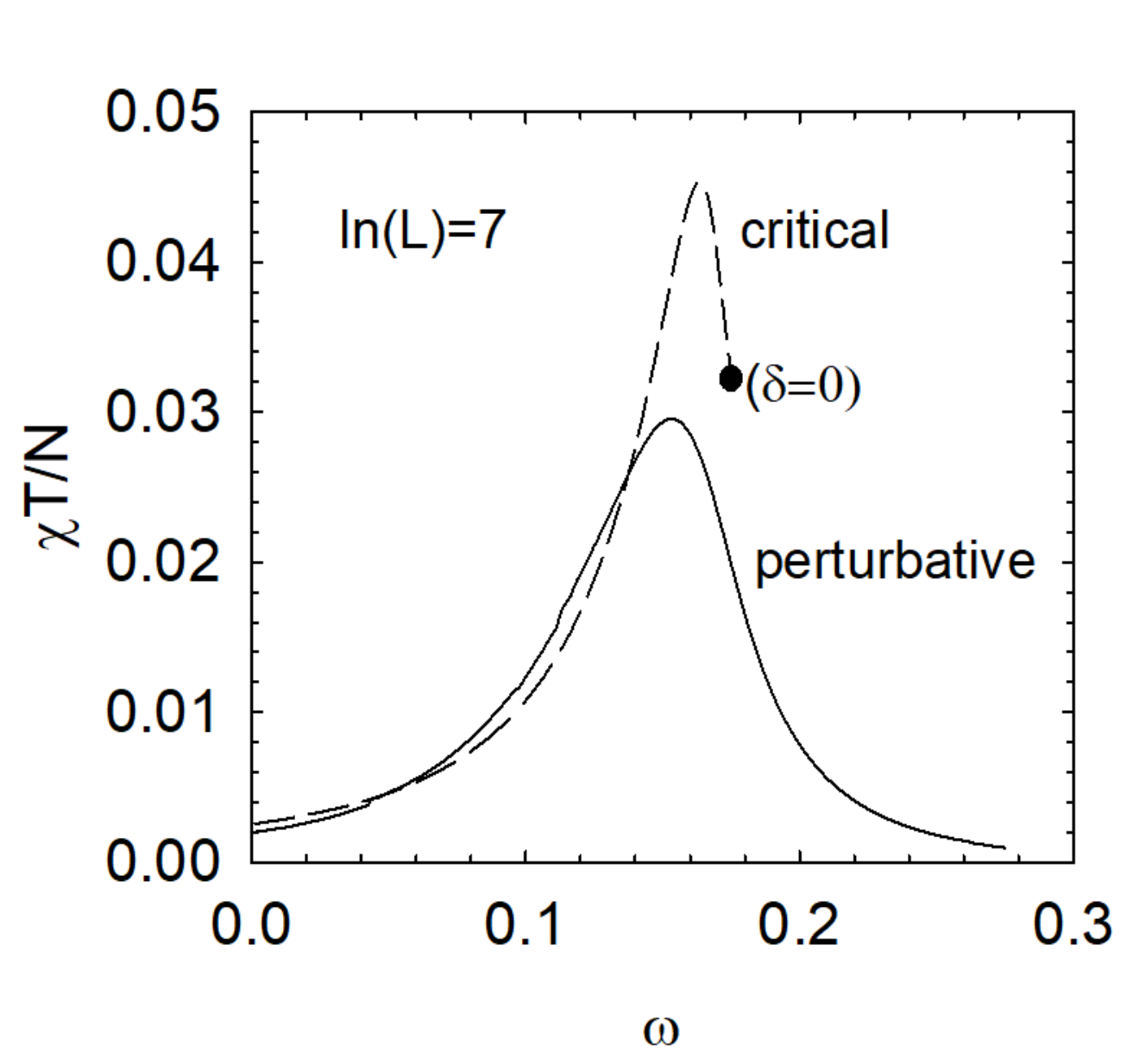}}
\caption{\label{compare} The calculated magnetic susceptibility is plotted agains $\omega$ using values of the effective exchange coupling $\delta(\omega)$ determined in the critical approximation (dashed line) and perturbative approximation (solid line) to the RG equations.  The curves are for a system size $\ln(L)=7$ and $D=1.9\times 10^{-5}$.  The solid dot indicates the point beyond which the exchange coupling is no longer ferromagnetic in the critical approximation, and the susceptibility calculation diverges. }
\end{figure}

A comparison of the magnetic susceptibility calculated using $\delta$ for a system with fourfold anisotropy, determined in both the critical and perturbative approximations in the previous sections,  is shown in fig.(\ref{compare}).  The system parameters are $\ln{L}=7$ and $D=1.9\times 10^{-5}$.   (A small jump in the solid curve near $\omega=0.11$ occurs at the point where the perturbative approximation for $\delta$ moves from one piecewise polynomial to another.)  Both curves are in qualitative agreement, in that the susceptibility is small in the spin wave region, begins to increase near $\omega_0 \approx 0.40$, where vortex pairs begin to form, and has a broad peak over the entire range of the finite-size transition.  

There are, however, important quantitative differences.  In the critical approximation, the susceptibility has a larger amplitude and reduced full-width at half maximum, and the position of the peak is significantly below $\omega_{c,L}\approx 0.18$. The curve terminates just past its peak, at the point where $\delta=0$ and eq.(\ref{chiS}) and (\ref{chiV}) diverge. Although the high temperature tail of the curve is predicted to have an exponential dependence on $\omega^{-1/2}$ from general arguments leading to eq.(\ref{xi}), it is not possible to demonstrate this characteristic functional dependence of the vortex gas.  In contrast, in the perturbative approximation $\delta \rightarrow 0$ asymptotically so that the system remains ferromagnetic, and the expressions  for the susceptibility remain well defined. The position of the curve maximum is very nearly at $\omega_L \approx 0.16$ and the form of the high temperature tail can be determined in detail.  These differences, and the changes in the values of $\omega_0$ and $\omega_L$, are important for quantitative fitting of experimental data to extract magnetic properties and properties of the vortex distribution.

\begin{figure}
\scalebox{0.5}{\includegraphics{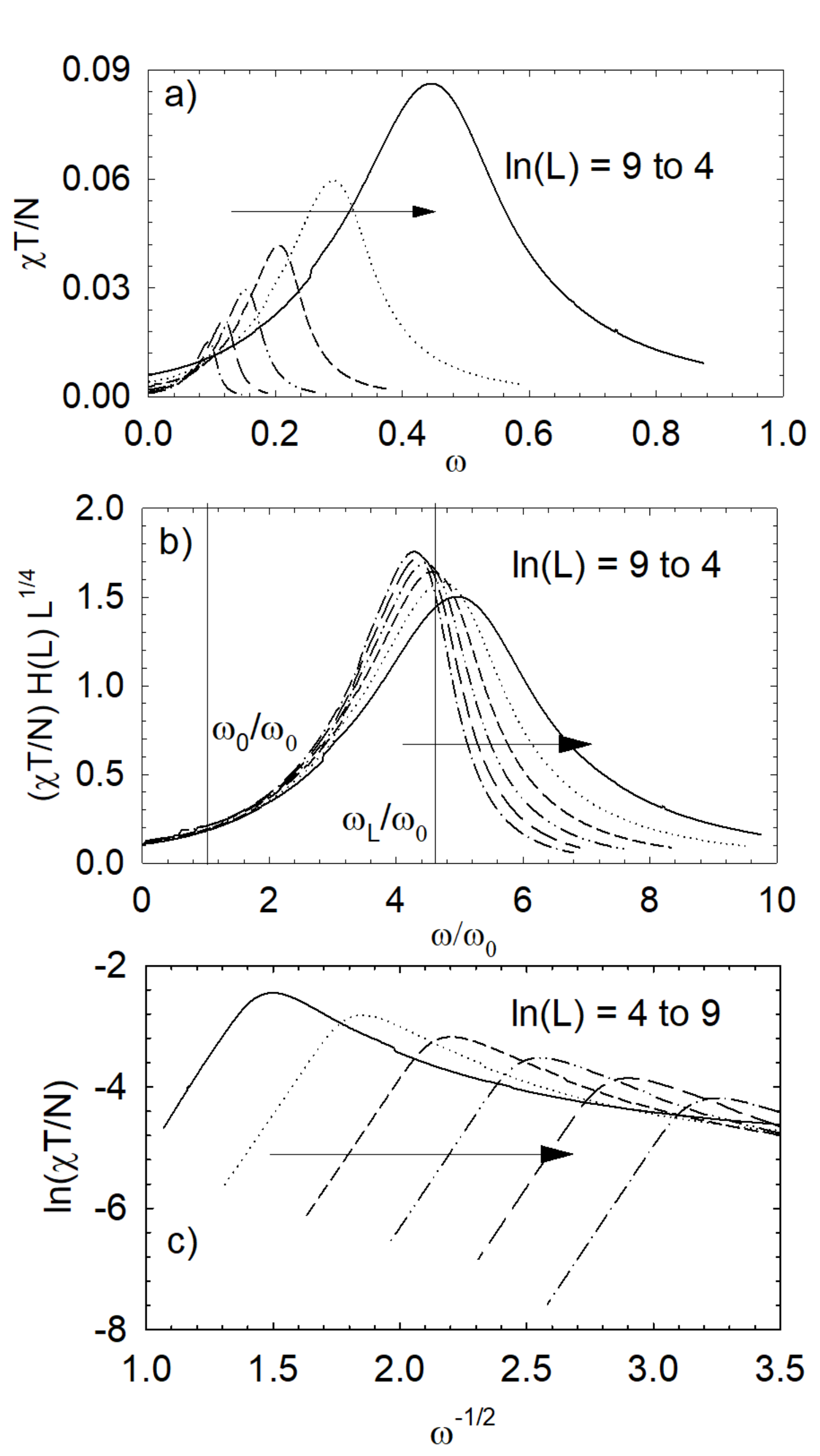}}
\caption{\label{chiplot} a) The magnetic susceptibility due to fluctuations in the magnitude of the magnetization (sum of eq.(\ref{chiS}) and (\ref{chiV})) is plotted against $\omega$ for different system sizes. The value $D=1.9\times 10^{-5}$ is used. b) The curves in part a) are replotted against $\omega/\omega_0$, and normalized by $H{(L)}\;L^{1/4}$ as defined in eq.(\ref{scaling}) to illustrate more clearly the change in the shape of the curve as the system size is changed.  c) The logarithm of the susceptibility at high temperature is plotted against $\omega^{-1/2}$ to illustrate the characteristic dependence of a vortex gas, as predicted by eq.(\ref{correl}). }
\end{figure}

For these reasons, further analysis of the magnetic susceptibility is restricted to that calculated using the perturbative approximation.  The sum of the spin and vortex contributions are shown in fig.(\ref{chiplot}a) for a range of system sizes.   It can be seen that the susceptibility gets narrower as the system size increases.  If it were not normalized by a factor of $N$ in the figure, the susceptibility per spin, $\chi$, would increase dramatically as the system size increased.  In the infinite, isotropic 2DXY model, the susceptibility $\chi(T_{KT})/N$ scales as $L^{-1/4} (\ln{L})^{1/8}$ at the KT transition.\cite{Kosterlitz1,Kosterlitz2} In a finite system, the transition begins instead at the reduced temperature $\omega_0$, and, as is shown in appendix C, the vortex susceptibility at this transition point scales as 
\begin{equation}
\chi_V(\omega_0) T/N \sim \omega_0 \ln{(\sqrt{b}L)}\; L^{-1/4}.
\end{equation}
In this expression, the explicit logarithmic term arises from essential finite size effects in the magnetization, and the factor of $\omega_0$ is due to finite-size effects in the coupling. As can be seen in fig.(\ref{width}a), the dependence of $\omega_0$ on $\ln{L}$ depends strongly on the approximations made in solving the RG equations.  In the critical approximation, the expression for $\omega_0$ in eq.(\ref{t0}) gives
\begin{equation}
\chi_V(\omega_0) T/N \sim (\ln{L})^{-1} L^{-1/4}.
\end{equation}
In the perturbative approximation, the dependence of $\omega_0$ on the size of the system can be expressed as a continued approximation in $\ln{L}$, using eq.(\ref{lnseries}).  This gives

\begin{equation}
\begin{split}
\chi_V(\omega_0) T/N &\sim \frac{L^{-1/4}}{[\ln{L} +\ln{(4\sqrt{e} \pi^{-1}\ln{L})}+ ...]}\:  \\
& \equiv \frac{1}{H(L)}  L^{-1/4},
\end{split}
\label{scaling}
\end{equation}
where $4\sqrt{e}/\pi \approx 2.1$.  This expression contains higher order logarithmic corrections.  In part (b) of the figure, the susceptibilities are scaled by $H(L)\L^{1/4}$, and plotted against $\omega/\omega_0$.  The data collapse at $\omega_0$ is very good. The fact that both the critical and perturbative approximations give the same order for the simple logarithmic correction provides some confidence that this order is correct.  The correction varying as $\sim \ln{[\ln{L}]}$ is necessary, as the scaling at $\omega_0$ is much better when it is included than when it is omitted. However, the exact order of this correction is likely sensitive to details of the perturbative approximation, such as the use of piecewise polynomials, and the truncation of the continued approximation for $\omega_0$ in orders of $\ln{L}$ after two terms.  The expression for the magnetic susceptibility itself has involved approximations in moving from eq.(\ref{lowT}) to eq.(\ref{chiV}).  The spin part of the susceptibility scales differently than the vortex part, but this does not effect the overall scaling because it is so much smaller (see appendix C).

This plot makes  it clear that the susceptibility peak becomes narrower in larger system sizes because the high temperature side is cut off more sharply.  This is due to the inclusion of larger vortices that more completely destroy the magnetization stabilized by finite-size effects.  It can also be seen that while the peak maxima occur near $\omega_L$, where $\delta(\omega)$ has a point of inflection, the peak position disperses somewhat with size.

The high temperature tail of the susceptibility is expected to scale\cite{Kosterlitz2} as $\chi \sim \xi^{2-\eta}$.  According to eq.({\ref{correl}}), it will therefore depend on reduced temperature as $\exp{[(2-\eta)(1.07\pi \omega^{-1/2})]}$, independent of system size.  This behaviour is illustrated in fig.(\ref{chiplot}c).  The slope of the curves ranges from 6.17 to 6.24 for system sizes of $\ln(L)$= 5 to 9, respectively.   If the predicted value of $\eta =1/4$ at $T_{KT}$ is used, then the slope is expected to be 5.88.  Because the slopes in fig.(\ref{chiplot}c) are determined significantly above $T_{KT}$, the value of $\eta$ is likely to be less than 1/4 and dependent on the temperature range.\cite{Nielson}  If this is indeed the case, then a value of $\eta = 0.16\pm 0.01$ is derived from the slopes.  

\subsection{Contribution due to fluctuations in the magnetization direction}
With the inclusion of explicit fourfold anisotropy, the direction of the magnetization may no longer be determined by the applied field, but rather by the magnetic easy axes.  It is then important to distinguish between the susceptibility with a small field applied along the magnetization (as in the previous section), and with a field applied perpendicular to the magnetization.  Experimental measurements are expected to include both.  

The anisotropy can be represented by an anisotropy field $\mathbf{H}^{an}$, and a small oscillating field $\mathbf{H}^{app}$ can be simultaneously parallel and perpendicular to an easy axis.  The effective field $\mathbf{H}^{eff}$ along which the scalar magnetization is aligned in equilibrium is
\begin{equation}
\mathbf{H}^{eff} = \mathbf{H}^{app} + \mathbf{H}^{an}.
\end{equation}
For definiteness, the x-axis is chosen along an easy axis, and the angle $\phi$ of the magnetization is measured from this axis.  Applying a field along the y-axis and measuring the magnetic response along the y-axis gives the measured susceptibility tensor component $\chi^{app}_{yy}$:
\begin{equation}
\label{recips}
\frac{1}{\chi^{app}_{yy}}=\frac{\partial H^{app}_y}{\partial M_y}=\frac{\partial H^{eff}_y}{\partial M_y} - \frac{\partial H^{an}_y}{\partial M_y},
\end{equation}
where the anisotropy field is derived from the anisotropy energy density\footnote{This section continues to use the same units as ref. 32.  For SI units factors of the saturation magnetization $M_S$ and magnetic permeability $\mu_0$ must be included. } $E^{an}$.
\begin{equation}
\label{Han}
H^{an}_y = -\frac{\partial E^{an}(M,\phi)}{\partial M_y}.
\end{equation}

Recalling that the effective field is by definition parallel to the magnetization in equilibrium, the reciprocal of the effective susceptibility component can be expressed in planar circular components as
\begin{equation}
\label{chieff}
\frac{\partial H^{eff}_y}{\partial M_y} =  \frac{\partial M}{\partial M_y} \frac{\partial }{\partial M} H^{eff} \sin{\phi}= \sin^2{\phi}\;  \frac{\partial H^{eff}}{\partial M}.
\end{equation}
$\partial H^{eff}/\partial M$ is just the (reciprocal of the) susceptibility due to fluctuations of the scalar magnetization calculated in the previous section for the finite-size KT transition.  For consistency of notation with previous sections, this susceptibility will be referred to simply as $\chi$.  Combining the results of eq.(\ref{recips}) to (\ref{chieff}), the experimentally measured susceptibility per spin is
\begin{equation}
\chi^{app}_{yy} = \frac{\chi}{\sin^2{\phi}+ \frac{\partial^2 E^{an}(M,\phi)}{ \partial M^2_y} \chi} .
\end{equation}

\begin{figure}
\scalebox{0.5}{\includegraphics{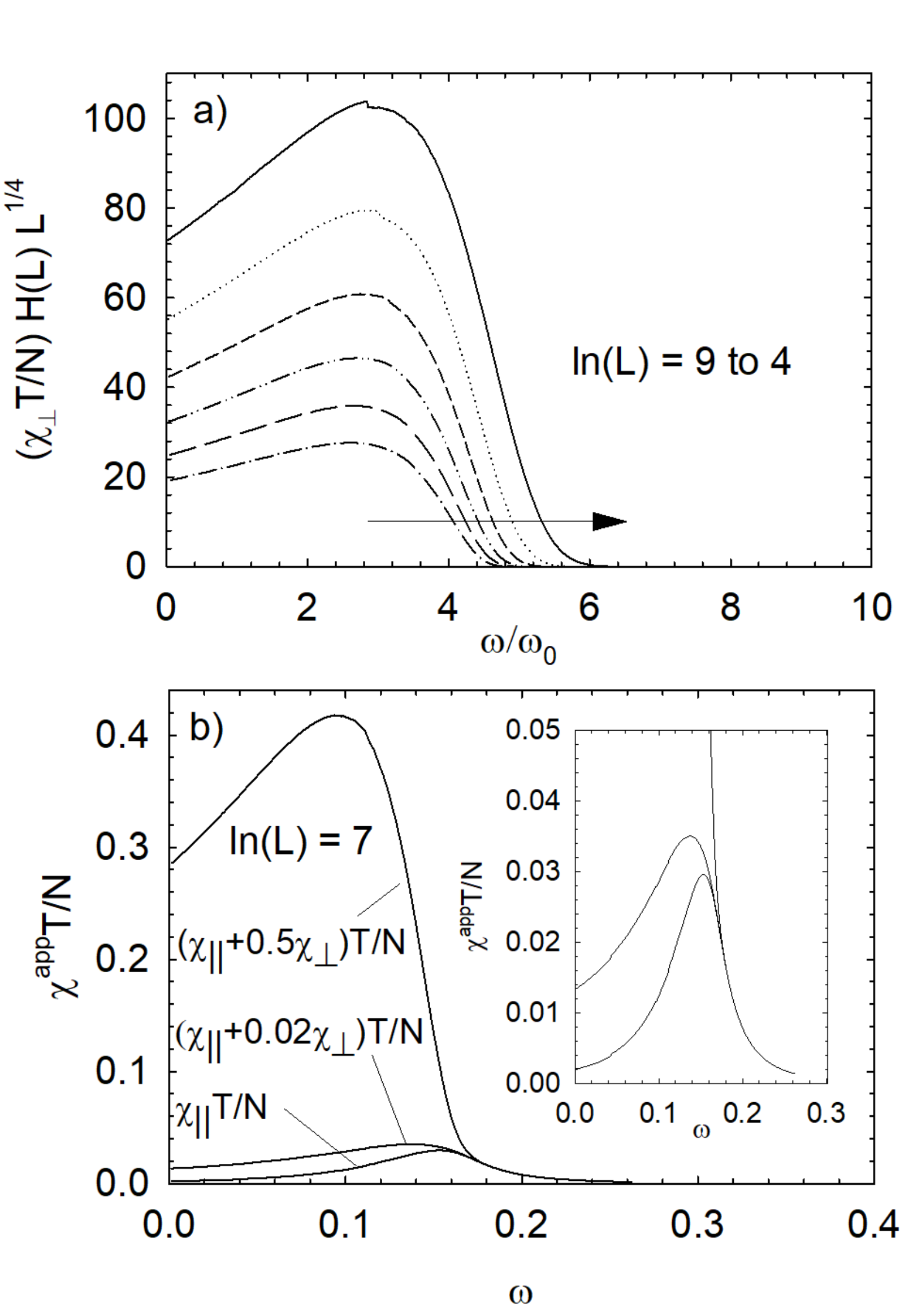}}
\caption{\label{chit} a) The transverse susceptibility in eq.(\ref{chiperp}) is plotted against $\omega/\omega_0$ for a range of system sizes. The susceptibility is scaled by $H(L)\;L^{1/4}$ as defined in eq.(\ref{scaling}), to allow direct comparison to the longitudinal susceptibility plotted in fig.(\ref{chiplot}b).  b) The measured susceptibility in an applied field, $\chi^{app}$, depends upon the alignment of the magnetization at low temperature and the applied field.  The curve labelled $\chi_{||}$ is for a low temperature domain aligned parallel to the applied field.  The curve with a small admixture of $\chi_\perp$ represents a low temperature domain slightly misaligned with the applied field.  The curve with a large admixture of $\chi_\perp$ represents a situtation where there is a distribution of low temperature domains aligned both parallel and perpendicular to the applied field.  The inset shows the high temperature region on an expanded scale.  Above $\omega \approx \omega_L$, the anisotropy disappears and $\chi_\perp \rightarrow 0$.  Then all the curves are given by $\chi_{||}$ and have the characteristic functional form of a KT transition.  The calculations use the parameters $\ln(L)=7$ and $D=1.9 \times 10^{-5}$.}
\end{figure}

Since the sum in the second term in eq.(\ref{Ham}) represents the total anisotropy energy of the system, dividing this sum by $N$ gives the anisotropy energy per spin. The sum itself is evaluated through the block spin renormalization procedure that is halted when the system is represented by a single block of size $L$, anisotropy parameter $Y_4(\ln{L})$ and spin orientation $\langle \theta \rangle =\phi$. The anisotropy energy per spin can therefore be written as\footnote{Again, in this section units with $k_B=1$ and lattice constants of unit length are used.}
\begin{equation}
E^{an} = \frac{2TY_4(\ln L)}{N} [1-\cos{4\phi}].
\end{equation}
Using planar circular co-ordinates once again to perform the partial derivatives in eq.(\ref{Han}) and (\ref{recips}) yields
\begin{equation}
\frac{\chi^{app}_{yy}T}{N} =  \frac{\chi T/N}{\sin^2{\phi} +\frac{32Y_4}{\langle M\rangle^2} [\chi T/N] \cos^2{\phi} \cos{4\phi}}.
\label{compchi}
\end{equation}
With the oscillating field applied along the y-axis, the low temperature domains with magnetization aligned along the easy axis parallel to the y-axis have $\phi=\pi/2$, and 
\begin{equation}
\label{chilong}
\frac{\chi^{app}_{yy}}{N}T \rightarrow \frac{\chi T}{N} \equiv \frac{\chi_{||} T}{N}.
\end{equation}
This is the result from the previous section.

For domains aligned along the easy axis parallel to the x-axis, $\phi = 0$ at low temperatures where the anisotropy persists, and the susceptibility is given by
\begin{equation}
\label{chiperp}
\frac{\chi^{app}_{yy} T}{N} \rightarrow  \frac{\langle M\rangle^2}{32Y_4} \equiv \frac{\chi_\perp T}{N}.
\end{equation}
As the temperature increases,  $Y_4$ decreases and goes to zero near $\omega_L$.  In the absence of anisotropy in eq.(\ref{compchi}), the scalar magnetization aligns with the applied field, giving $\phi = \pi/2$.  Then the susceptibility is once again given by eq.(\ref{chilong}).  For a sample with a distribution of both domain types, the susceptibility will be given by a linear combination of the limiting forms $\chi_{||}$ and $\chi_\perp$.

The transverse susceptibility $\chi_\perp T/N$ in eq.(\ref{chiperp}) is plotted in fig.(\ref{chit}a) for a range of system sizes. The susceptibility is scaled in the same way as the plot of $\chi_{||} T/N$ in fig.(\ref{chiplot}b) to allow comparison.  It can be seen that the transverse susceptibility due to angular fluctuations about the easy axis is much larger than the longitudinal susceptibility due to fluctuations of the magnitude of the magnetization. In addition, the low temperature limit (near $\omega=0$) of $\chi_\perp$ increases much more quickly as the system size is decreased, than does $\chi_{||}$.  However, as the anisotropy is screened near and above $\omega_L$, $\chi_\perp T/N$ goes quickly to zero and the longitudinal susceptibility is dominant. As a result, the high temperature tail of the susceptibility displays the characteristic functional form of a KT transition regardless of the domain orientations at low temperature.  This is consistent with an isotropic paramagnetic vortex gas.   

This behaviour can be seen in part (b) of the figure.  The curve labeled $\chi_{||}T/N$ represents a low temperature domain with the magnetization aligned with the applied field, whereas that with a small admixture of $\chi_\perp$ represents a situation where the field direction is slightly misaligned with the magnetization.  These two curves are shown more clearly in the insert to the figure, using a magnified scale.  These curves look very much like the experimental curves categorized as Type I in the experimental investigation of Fe/W(001) films by Atchison \textit{et al.}\cite{Atchison}  The third curve, with a large admixture of of $\chi_\perp $, represents a situation where there are equal portions of low temperature domains aligned along each of the two easy axes.  This curve is similar to those categorized as Type II in the experimental study, including the observed factor of roughly ten in amplitude compared to Type I measurements.  Although the precise numerical factors for the admixtures in fig.(\ref{chit}b) are not fitted, but rather chosen for illustrative purposes, the similarity between these first-principles calculations and the experimental measurements is very encouraging.  These results support the suggestion that the difference between Type I and Type II measurements in the finite-size KT transition has to do with the low temperature domain distribution in the film and the distinction between $\chi_\perp$ and $\chi_{||}$ introduced by the fourfold anisotropy.

\section{Conclusions}
The magnetic response of the \textit{finite, anisotropic} 2DXY model has been investigated using the renormalization group equations, by extending previous work on the \textit{infinite, anisostropic} model and the \textit{finite, isotropic} model.  An exact solution of the RG equations in the critical approximation confirms previous numerical simulations that showed that a finite-size KT transition is preserved for low anisotropy, and a 2D Ising transition occurs at high anisotropy.  The boundary line between these two behaviours depends upon the product of the anisotropy and system size through the system parameter $B=2\gamma \sqrt{D} \ln{L}$.  In a more restricted region of parameter space, the effect of fourfold anisotropy can be incorporated into the description of the $\textit{isotropic}$ system by the simple substitution $\alpha t \to \omega \equiv \alpha t + 2\gamma^2 D$. The temperature range within which the critical approximation is valid is considerably narrower than the finite-size KT transition itself, so that the coupling, fourfold anisotropy and magnetic susceptibility found in this way provide qualitative insight, but are not quantitatively reliable. 

The results of the critical approximation validate solving the RG equations by treating the fourfold anisotropy as a perturbation.  This approximation proves to be quantitatively reliable across the full temperature width of the finite-size KT transition, so long as the system parameter $B$ is small enough.  A principle finding is that the coupling no longer exhibits physically unreasonable behaviour (moving from ferromagnetic to antiferromagnetic, and then diverging), but rather approaches zero asymptotically.  The universal jump of the coupling seen in an infinite system becomes instead a universal  inflection point of steepest descent where $\omega_L = 4.61 \omega_0$ and $\delta = 0.395$.  At $\omega_L$, the unbinding of vortex-antivortex pairs becomes significant.  The dependence of $\omega_L$ on system size implies that the correlation length and magnetic susceptibility retain the exponential temperature dependence characteristic of the KT transition. Furthermore, the fourfold anisotropy $Y_4(\ell, \omega)$ calculated in the perturbative approximation no longer exhibits an unphysical cusp, or becomes complex, near $\omega_L$

The perturbative approximation gives an improved calculation of the effective exchange coupling, vortex density and anisotropy throughout the $\sim 10K$ range of the finite-size KT transition in a ferromagnetic film with fourfold anisotropy. These  in turn permit an improved calculation of the magnetic susceptibility, $\chi_{||}$, due to the fluctuations in the magnitude of the magnetization.  The improved results for the anisotropy and scalar magnetization are used to find the transverse susceptibility, $\chi_\perp$, for angular fluctuations of the magnetization about an easy axis.  Together, these susceptibility components give a more complete account of the magnetic response of the 2DXY model with fourfold anisotropy.

An initial comparison to the measurements of the magnetic susceptibility of Fe/W(001) ultrathin films is very encouraging.  Suitable combinations of $\chi_{||}$ and $\chi_\perp$ are in good qualitative agreement with, for instance, fig. (1a) and (4a) in Atchison \textit{et al.}\cite{Atchison}. In particular, the two distinct shapes of the measured susceptibility termed Type I and Type II by those authors agree well with the combinations of  $\chi_{||}$ and $\chi_\perp$ expected for situations where low temperature magnetic domains are aligned along single or multiple fourfold easy axes.  Because $\chi_\perp \rightarrow 0$ near $\omega_L$, both types of measurements exhibit the exponential dependence on temperature characteristic of a finite-size KT transition, indicating an isotropic high temperature phase. The experimental value of $\eta = 0.12\pm 0.09$ in the temperature range where the exponential dependence is observed is consistent with the present calculations, where $\eta = 0.16\pm 0.01$.  The implication is that careful fitting of Type I measurements can be used to understand details of vortex pair formation, and fitting of Type II measurements can be used to study the evolution of the anisotropy in the finite-size KT transition.  This process is underway.

These results open numerous opportunities to study spin wave and vortex properties in an ultrathin ferromagnetic film.  The RG treatment of the 2DXY model uses an effective medium approach, where the presence of vortices and bound vortex pairs alters the medium in which spin waves propagate.  Therefore, these calculations can be a basis for interpreting the imaginary, dissipative components of the measured susceptibility.  For example, $Y_4(\ell, \omega)$ can be used to determine the energy barrier to dissipative domain switching due to an applied field.  Also, the variation of the domain wall energy and activation energy for domain wall pinning in the effective medium are determined by a combination of $Y_4(\ell, \omega)$ and $\delta(\ell, \omega)$.  These dissipative process can be studied as the system moves from  a low temperature system dominated by spin wave excitations to one dominated by vortices.  The imaginary component of the susceptibility above the transition may provide information on the dynamics of the vortex gas itself.   These investigations are underway.

\renewcommand{\theequation}{A.\arabic{equation}}
\setcounter{equation}{0}
\section*{Appendix A. Exact solution in the critical approximation}

Beginning with eq.(\ref{finite})
\begin{equation}
-\int d\ell = \int_{x_i}^{x_f} \frac{dx}{\sqrt{x^2+\alpha t}\;\sqrt{x^2+\alpha t+4\gamma^2 D}},
\end{equation}
a standard transformation will show that this is an elliptic integral of the first kind.  Since the flow is from a larger positive initial value of $x_i$ to a final value near $x_f =0$, it is advantageous to write the integral in a form where it is dominated by the endpoint near $x_f =0$.  Then it is insensitive to the initial value and the scaling properties will not depend it, as is expected. This can be accomplished through the co-ordinate transformation $z=1/x$. Then
\begin{equation}
-\int d\ell = \int_{1/x_i}^{1/x_f} \frac{-dz}{\sqrt{1+\alpha t z^2}\;\sqrt{1+(\alpha t+4\gamma^2 D)z^2}}.
\end{equation}
Letting $\nu = \sqrt{\alpha t +4\gamma^2 D}\: z$,
\begin{equation}
\sqrt{\alpha t +4\gamma^2 D} \int d\ell = \int_{\frac{\sqrt{\alpha t +4\gamma^2 D}}{x_i}}^{\frac{\sqrt{\alpha t +4\gamma^2 D}}{x_f}}
\frac{d\nu}{\sqrt{(1+\nu^2)(1+(k')^2\nu^2)}},
\end{equation}
where $(k')^2 = (\alpha t)/(\alpha t +4\gamma^2 D)$.  This is a well-known transformation of the standard form of elliptic integrals of the first kind\cite{handbook}, obtained by letting $\nu = \tan\phi$. Then
\begin{equation}
\sqrt{\alpha t +4\gamma^2 D} \int d\ell = \int_{\arctan(\frac{\sqrt{\alpha t +4\gamma^2 D}}{x_i})}^{\arctan(\frac{\sqrt{\alpha t +4\gamma^2 D}}{x_f})}
\frac{d\phi}{\sqrt{1- k^2 \sin^2 \phi}},
\label{elliptic}
\end{equation}
where $k^2 = 1- (k')^2 =(4\gamma^2 D)/(\alpha t +4\gamma^2 D)$.  At this point, as the value of $x_i$ is not important, let $x_i \rightarrow \infty$.
Recalling that the upper limit of $\ell$ is $\ln L$ (in units of the lattice constant), the integral becomes
\begin{equation}
\sqrt{\alpha t +4\gamma^2 D}\: \ln L = F(\phi_f, k),
\label{1stkind}
\end{equation}
with $F(\phi_f, k)$ the elliptic integral of the first kind, $\phi_f = \arctan(\sqrt{\alpha t +4\gamma^2 D}/x_f )$, and $x_f=x(\ln{L})$.

The expression in eq.(\ref{1stkind}) can be formally written in a way that isolates $x(\ln{L})$ by using the inverse elliptic function\cite{handbook} $\text{am} (u,k)$.  If
\begin{equation}
u=F(\phi_f,k), 
\label{defam}
\end{equation}
then the inverse function is defined as
\begin{equation}
\phi_f=\text{am}(u,k).
\end{equation}
In the present case,
\begin{equation}
\arctan(\sqrt{\alpha t +4\gamma^2 D}/x(\ln{L})) = \text{am}(\sqrt{\alpha t +4\gamma^2 D}\: \ln L, k),
\end{equation}
\begin{equation}
x(\ln{L}) = \frac{\sqrt{\alpha t +4\gamma^2 D}}{\tan[\text{am}(\sqrt{\alpha t +4\gamma^2 D}\: \ln L, k)]}.
\end{equation}

Although the inverse elliptic integral cannot be solved analytically, some general results can be extracted by employing the Jacobi elliptic functions $\text{sn}(u,k), \text{and cn}(u,k)$, defined as\cite{handbook}
\begin{equation}
\text{sn}(u,k) = \sin[\text{am}(u,k)] = \sin \phi_f,
\end{equation}
\begin{equation}
\text{cn}(u,k) = \cos[\text{am}(u,k)] = \cos \phi_f.
\label{cn}
\end{equation}
First note that, for the present problem, 
\begin{equation}
\begin{split}
uk &= (\sqrt{\alpha t +4\gamma^2 D}\: \ln L) \sqrt{\frac{4\gamma^2 D}{\alpha t +4\gamma^2 D}}\\
&= \sqrt{4\gamma^2 D}\: \ln L \equiv B,
\end{split}
\label{defB}
\end{equation}
where $B$ is a constant for a given system of size $L$ and anisotropy $h_4 \rightarrow \sqrt{D}$.
Then the expression for $x(\ln{L})$ can be written as
\begin{equation}
x(\ln{L}) = \frac{\sqrt{4\gamma^2 D}}{k} \: \frac{\text{cn}(B/k , k )}{\text{sn}(B/k,k)}.
\label{final}
\end{equation}

To determine the functional form of $Y_4(\ln{L})$ and $Y(\ln{L})$ in this temperature range, eq.(\ref{sep}) can be combined with eq.(\ref{final}) to give
\begin{equation}
x^2=\frac{4\gamma^2 D}{k^2}\frac{\text{cn}^2(B/k,k)}{\text{sn}^2(B/k,k)}=\gamma^2(\frac{D}{Y_4}-Y_4)^2-\alpha t.
\end{equation}
Using the property\cite{handbook} $\text{cn}^2(u,k)=1-\text{sn}^2(u,k)$, and reversing the sign of the square in $Y_4(\ln{L})$,
\begin{equation}
\begin{split}
\frac{4\gamma^2 D}{k^2\text{sn}^2(B/k,k)} & - \frac{4\gamma^2 D}{k^2} \\
&=\gamma^2 (\frac{D}{Y_4}+Y_4)^2-4\gamma^2 D-\alpha t.
\end{split}
\end{equation}
Recalling the definition of $k^2$, the second term on the left side cancels with the last two terms on the right.  The third Jacobi elliptic function\cite{handbook},
\begin{equation}
\text{dn}(u,k)=\sqrt{1- k^2\:\text{sn}^2(u,k)},
\end{equation}
is used to substitute for $\text{sn}^2(B/k,k)$.  Then the square root of both sides can be taken to give
\begin{equation}
\frac{2\sqrt{D}}{\sqrt{1-\text{dn}^2(B/k,k)}}=\frac{D}{Y_4}+Y_4.
\end{equation}
This quadratic equation, and the corresponding quadratic equation for $Y$, can be solved for two physical roots:
\begin{equation}
\frac{Y_4}{\sqrt{D}}=\sqrt{\frac{1-\text{dn}(B/k,k)}{1+\text{dn}(B/k,k)}},
\label{Y4ofk}
\end{equation}
\begin{equation}
\frac{Y}{\sqrt{D}}=\sqrt{\frac{1+\text{dn}(B/k,k)}{1-\text{dn}(B/k,k)}}.
\label{Yofk}
\end{equation}

The condition for the beginning of the finite-size transition\cite{Bramwell1}  is that $K=2/\pi$, or $x=0$ .  According to eq.(\ref{final}), this corresponds to 
\begin{equation}
\text{cn}(B/k_0,k_0)=0,
\end{equation}
where \begin{equation}
k_0=\sqrt{\frac{4\gamma^2 D}{\alpha t_0 +4\gamma^2 D}}
\end{equation}
defines the value of reduced temperature $t_0$ where the transition begins.  The periodic property\cite{handbook} of the  function $\text{cn}(u,k)$ requires that
\begin{equation}
 B/k_0 = \kappa(k_0),
 \label{k0}
\end{equation}
with $\kappa(k_0)$ the \textit{complete} elliptic integral of the first kind. Combining the definition of $B$ in eq.(\ref{defB}) with these results gives an expression for the transition point.
\begin{equation}
\alpha t_0 = (1-k_0^2)\;[\frac{\kappa (k_0)}{\ln L}]^2.
\end{equation}

In the limit of vanishing fourfold anisotropy, $k_0 \rightarrow 0$, and $\kappa(0)=\pi/2$, so that finite-size transition point is in agreement with eq.(\ref{t0}).  The end of the finite-size transition, where the vortices unbind, occurs when\cite{Bramwell1} $x\rightarrow -\infty$ at $k=k_L $.  This condition is given by eq.(\ref{final}) as
\begin{equation}
\text{sn}(B/k_L, k_L)=0,  \: \text{or} \: B/k_L = 2\kappa(k_L).
\end{equation}
This implies $k_L \le k_0$.  In the limit where there is no anisotropy, this agrees with eq.(\ref{tL}).  The correlation length at and above $t_L$ is
\begin{equation}
\xi \sim L=\exp{[\frac{2\kappa (k_L)}{\sqrt{\alpha t + 4\gamma^2 D}}]}; \:\: t > t_L,
\end{equation}
and has the functional form expected for a KT transition.

\renewcommand{\theequation}{B.\arabic{equation}}
\setcounter{equation}{0}
\section*{Appendix B. Polynomial approximation of the integral expression for $\delta(\ell)$}

The critical approximation uses the quadratic approximation $f(\delta)=1/2\: (\delta-1)^2$.  When this expression is used in eq.(\ref{dfinite}) of the perturbative approximation, the integral for $\delta(\ell)$ become
\begin{equation}
\frac{d\delta}{4[\frac{1}{2}(1-\delta)^2 ]+\frac{1}{2}\omega \delta^2} = \frac{d\delta}{X(\delta)} = -d\ell,
\label{dsmallx}
\end{equation} 
where
\begin{equation}
X(\delta) = a +b\delta + c\delta^2.
\end{equation}
with $a=2,\:b=-4$ and $c=2+\frac{\omega}{2}$.  This standard integral yields the expression\cite{handbook}
\begin{equation}
\delta=\frac{1+\frac{\sqrt{\omega}}{2\tan{[\sqrt{\omega}\ln L]}}}{1+\frac{\omega}{4}}.
\label{quadratic}
\end{equation}
For comparison, the solution for the critical approximation, given by eq.(\ref{xapprox}), is
\begin{equation}
\delta=1+\frac{\sqrt{\omega}}{2\tan{[\sqrt{\omega}\ln L]}}.
\label{nearx=0}
\end{equation}

A better representation of $f(\delta)$ is given by two piecewise polynomials.  For $\delta \ge 3/4$, the integral expression for $\delta(\ell)$ in eq.(\ref{dfinite}) is well-approximated by
\begin{equation}
\frac{d\delta}{4[\frac{1}{2} \delta (1-\delta)^2 ]+\frac{1}{2}\omega \delta^2} = -d\ell.
\label{Alarged}
\end{equation}
The integral in $\delta$ is now of the form
\begin{equation}
\int_{\delta_i}^{\delta_f} \frac{d\delta}{\delta X(\delta)}
\label{Aint}
\end{equation}
with $a=c=2$, and $b=-4 +\omega /2$, and the discrimanent $q=4ac-b^2=\frac{1}{2}\omega (8 - \frac{1}{2}\omega)$.
The solution is\cite{handbook}
\begin{equation}
\frac{1}{2a}\ln \frac{\delta^2}{X(\delta)}|_{\delta_i}^{\delta_f} - \frac{b}{a\sqrt{q}}\arctan{[\frac{2c\delta +b}{\sqrt{q}}]}|_{\delta_i}^{\delta_f},
\end{equation}
Evaluating this expression in the limit $\delta_i \rightarrow \infty$ is well-behaved, and gives
\begin{equation}
\frac{1}{2a}\ln c + \frac{b}{a\sqrt{q}} \frac{\pi}{2}.
\end{equation}
After considerable algebra, the following closed expression for $\delta_f = \delta(\ell)$ is obtained:
\begin{equation}
\delta = 1-\frac{\omega}{8}+\frac{\sqrt{\frac{\omega}{2}(8-\frac{\omega}{2})}}{4\tan{[\frac{\sqrt{\frac{\omega}{2}(8-\frac{\omega}{2})}}{2-\frac{\omega}{4}} \ln{[L\;(\frac{\delta^2}{(1-\delta)^2+\frac{\omega\delta}{4}})^{\frac{1}{4}}]}]}}.
\label{exactd}
\end{equation}
In the limit that $\omega /8 \ll 1$, this reduces to eq.(\ref{larged2}).

When $\delta=1$, the coupling reaches the critical value where the finite-size KT transition begins.  This is denoted by $\omega =\omega_0$.  This condition can be found by using a small angle approximation for the tangent function in eq(\ref{exactd}), when the angle is just less than $\pi /2$.  Then
\begin{equation}
\frac{\pi}{2\sqrt{\omega_0}} [\frac{1-\frac{\omega_0}{8}}{\sqrt{1-\frac{\omega_0}{16}}}] - \frac{1}{4}[\frac{1-\frac{\omega_0}{8}}{1-\frac{\omega_0}{16}}]= \ln{[L(\frac{4}{\omega_0})^{1/4}]}.
\label{exactw0}
\end{equation}
 In the limit where $\omega_0 /8 \ll 1$, this reduces to eq.(\ref{lnseries}).

For $\delta \le 3/4$, the polynomial approximation to $f(\delta)$ in fig.(\ref{f(delta)}) is given by
\begin{equation}
\delta[\delta^2-1.73\delta+0.77)]=\delta[(\frac{7}{8}-\delta)^2]+\frac{1}{40}\delta^2.
\end{equation}
This leads to an integral for $\delta(\ell)$ of the same form as eq.(\ref{Aint}), but with $a=(7/4)^2,\:b=-7+(1/10 +\omega/2),\:c=4$ and 
\begin{equation}
q=(\frac{1}{10}+\frac{\omega}{2})(14-(\frac{1}{10}+\frac{\omega}{2})).
\end{equation}
After considerably more algebra, the result is
\begin{equation}
\begin{split}
&\delta=  \frac{6.9-\frac{\omega}{2}}{8} +\\
 &\frac{\sqrt{(\frac{1}{10}+\frac{\omega}{2})(13.9-\frac{\omega}{2})}}{8\tan{[\frac{49}{16}\frac{ \sqrt{(\frac{1}{10}+\frac{\omega}{2})(13.9-\frac{\omega}{2})}}{(6.9-\frac{\omega}{2})}[\ln{L(\omega)+\frac{8}{49}\ln(\frac{\delta^2}{(\frac{7}{4}-2\delta)^2+(\frac{1}{10}+\frac{\omega}{2})\delta})}]]}}.
\end{split}
\label{smalld}
\end{equation}
In this expression,
\begin{equation}
\begin{split}
&\ln L(\omega)=  \ln(L) +\frac{1}{4}\ln[\frac{9}{1+3\omega}]-\frac{8}{49}\ln[\frac{9}{2(1+3\omega)+\frac{1}{5}}]\\
& -\frac{(2-\frac{\omega}{4})}{\sqrt{\frac{\omega}{2}(8-\frac{\omega}{2})}}\arctan{[\frac{-\sqrt{\frac{\omega}{2}(8-\frac{\omega}{2})}}{1-\frac{\omega}{2}}]} \\
& +\frac{16}{49}\frac{(6.9-\frac{\omega}{2})}{\sqrt{(\frac{1}{10}+\frac{\omega}{2})(13.9-\frac{\omega}{2})}}\arctan{[-\frac{\sqrt{(\frac{1}{10}+\frac{\omega}{2})(13.9-\frac{\omega}{2})}}{0.9-\frac{\omega}{2}}]}.
\end{split}
\label{L(w)}
\end{equation}
The complicated expression for $\ln L(\omega)$ arises due to matching the two quadratic approximations at $\delta=3/4$.  Note that both of the arctangent functions return angles in the 2nd quadrant.

\renewcommand{\theequation}{C.\arabic{equation}}
\setcounter{equation}{0}
\section*{Appendix C. Evaluation of the susceptibility}
In the low temperature, spin wave limit of the harmonic model, Archambault \textit{et al.}\cite{Archambault1} show that the magnetization is of the form
\begin{equation}
\langle M \rangle= \exp{(-\frac{G(0)}{2K})},
\end{equation}
where $G(0)$ is the Green's function propagator for the square lattice, evaluated at the origin.  That is,
\begin{equation}
G(r)=\frac{1}{N}\sum_{q\ne 0} \frac{e^{-iq\cdot r}}{\epsilon_q},
\end{equation}
evaluated at $r=0$. In this Fourier sum over wavevectors $q$ in 2D,
\begin{equation}
\epsilon_q = 4-2\cos{q_x}-2\cos{q_y}.
\end{equation}
A discrete evaluation gives
\begin{equation}
\label{G0}
G(0)=\frac{\ln{(bN)}}{4\pi},
\end{equation}
where\cite{Bramwell5} $b=1.845...$, as before. Because of this logarithmic dependence on system size, the 2DXY model has intrinsic finite-size effects, with
\begin{equation}
\langle M \rangle=(\frac{1}{bN})^{\frac{1}{8\pi K}}
\end{equation}
converging very slowly even for macroscopic $N$.

\subsection{The vortex susceptibility}

Since the experimental system\cite{Atchison} has $\ln{L} \approx 7$, the continuum limit of the sum should be a very good approximation.  Because $\delta$ is a function of the scalar $\ln{L}$ in the perturbative approximation, there is no differentiation between the in-plane $x$ and $y$ axes, and the integral can be most easily performed in circular, planar co-ordinates $(\rho, \theta)$.  In moving from a square to a circular system while maintaining the number of spins,
\begin{equation}
N = L^2 = \frac{\pi}{4} \Delta^2,
\end{equation}
where $\Delta/2=L/\sqrt{\pi}$ is the maximum value of $\rho$.  The minimum value of $\rho$, corresponding to the bare lattice spacing before geometric scaling, is $1/\sqrt{\pi}$.  In the continuum limit, the vortex susceptibility in eq.(\ref{chiV}) is
\begin{equation}
\begin{split}
\frac{\chi_V T}{N} =& \frac{1}{N} [\int_0 ^{2\pi} d\theta \int_\frac{1}{\sqrt{\pi}} ^{\frac{\Delta}{2}} \rho d\rho \exp{(-\frac{\pi G(0)}{2\delta(\ln{2\rho})})}] \\
& -\exp{(-\frac{\pi G(0)}{2\delta(\ln{\Delta})})}.
\end{split}
\label{vortexint}
\end{equation}
Using the change of variables $\ell = \ln{2\rho}$,
\begin{equation}
\begin{split}
\frac{\chi_V T}{N} =& \frac{\pi}{2N} [ \int_{\ln{\frac{2}{\sqrt{\pi}}}} ^{\ln{\Delta}} d\ell \exp{(2\ell)} \exp{(-\frac{\pi G(0)}{2\delta(\ell)})}] \\
& -\exp{(-\frac{\pi G(0)}{2\delta(\ln{\Delta})})}.
\end{split}
\end{equation}
Finally, substituting for $G(0)$ and $N$,
\begin{equation}
\begin{split}
\frac{\chi_V T}{N} = &\frac{2}{\Delta^2} [ \int_{\ln{\frac{2}{\sqrt{\pi}}}} ^{\ln{\Delta}} d\ell \exp{(2\ell)} \exp{(-\frac{\ln{(\sqrt{b\pi}\Delta/2)}}{4\delta(\ell)})}] \\
& -\exp{(-\frac{\ln{(\sqrt{b\pi}\Delta/2)}}{4\delta(\ln{\Delta})})}.
\end{split}
\end{equation}
Recall that this is an equation for $\chi_V(\omega)$ because of the implicit temperature dependence of $\delta(\ell,\omega)$.  In the low temperature limit, the coupling $\delta$ renormalizes very slowly with size, so that it is essentially constant.  Then the vortex susceptibility is identically zero.

To display the scaling properties of the vortex susceptibility, the first term in eq.(\ref{vortexint}) is integrated by parts. One portion of the integration by parts cancel exactly with the second term in eq.(\ref{vortexint}), leaving
\begin{equation}
\frac{\chi_V(\omega) T}{N} = -\frac{8}{\Delta^2} \int_\frac{1}{\sqrt{\pi}} ^{\frac{\Delta}{2}} d\rho \;\rho^2 \frac{ d\delta}{d\rho} \frac{\pi G(0)}{2\delta^2(\rho,\omega)}  \exp{(-\frac{\pi G(0)}{2\delta(\rho,\omega)})}.
\label{vdu}
\end{equation}
Here the dependence of the susceptibility on $\omega$ through the coupling is displayed explicitly.  The universal point where the finite-size transition begins, regardless of the system size, is $\omega =\omega_0$, where $\delta =1.$  In the finite, anisotropic system, eq.(\ref{exactd}) is used  to give
\begin{equation}
\frac{d\delta(\rho,\omega_0)}{d\rho} |_{\delta\rightarrow1} = -\frac{\omega_0}{2\rho} \: \frac{1}{1+ \sin^2{[\sqrt{\omega_0} \ln{[\rho (\frac{4}{\omega_0})^{1/4}]]}}}.
\end{equation}
Eq.(\ref{lnseries}) is used to substitute for the expression $\ln{(4/\omega_0)^{1/4}}$ within the argument of the sine function.  
\begin{equation}
\frac{d\delta(\rho,\omega_0)}{d\rho} |_{\delta\rightarrow1} = -\frac{\omega_0}{2\rho} \frac{1}{1+ \sin^2{[\frac{\pi}{2}+\sqrt{\omega_0}[\ln{(\rho /\Delta) -1/4}]]}}.
\end{equation}
Because $\omega_0$ is small, the sine function is essentially unity.  Replacing these results, and the expression for $G(0)$ from eq.(\ref{G0}), into eq.(\ref{vdu}) produces
\begin{equation}
\begin{split}
\frac{\chi_V(\omega_0) T}{N} \approx &\frac{ \omega_0}{2 \Delta^2} \ln{(\sqrt{b\pi}\Delta/2}) \times \\
&\int_\frac{1}{\sqrt{\pi}} ^{\frac{\Delta}{2}} \frac{d\rho \;\rho }{ \delta^2(\rho,\omega_0)} (\sqrt{b\pi}\Delta/2)^{-\frac{1}{4\delta(\rho,\omega_0)}}.
\end{split}
\end{equation}
This integral contains only powers of  $1/\delta(\rho,\omega_0)$, and this is well-behaved near $\omega_0$, as can be seen in fig.(\ref{delta}). The integral over $\rho\;d\rho$ will cancel the prefactor of $\Delta^{-2}$. The expected scaling with size at the onset of the finite-size transition is therefore approximately
\begin{equation}
\frac{\chi_V(\omega_0) T}{N} = \frac{4\chi_V(\omega_0) T}{\pi \Delta^2} \sim \omega_0 \ln{(\sqrt{b\pi}\Delta/2}) \;\Delta^{-1/4}.
\end{equation}
The detailed scaling of the susceptibility depends on the dependence of $\omega_0$ on $\ln{L}$. As can be seen in fig.(\ref{width}a), this depends upon the approximations made in the solution of the RG equations.

\subsection{The spin wave susceptibility}
In the continuum limit, the expression for the spin wave susceptibility in eq.(\ref{chiS}) can be written as
\begin{equation}
\begin{split}
\frac{\chi_S T}{N} = &\frac{\pi^2}{8N} \int_{\frac{1}{\sqrt{\pi}}} ^{\frac{\Delta}{2}}    \frac{d\rho}{\delta^2(\ln{2\rho})} \exp{(-\frac{\pi G(0)}{2\delta(\ln{2\rho})})}\\
& \times \int_0 ^{2\pi} d\theta \rho\; G^2(\rho,\theta).
\end{split}
\end{equation}
This can be integrated by parts by identifying
\begin{equation}
dv =  \int_0 ^{2\pi} d\theta \rho\; G^2(\rho,\theta) d\rho,
\end{equation}
so that 
\begin{equation}
v = \int d\rho \int_0^{2\pi} d\theta \rho\; G^2(\rho,\theta) \equiv \sum_{r=1}^{\aleph(\rho)} G^2(r).
\end{equation}
In this expression, $\aleph(\rho)=\pi \rho^2$ limits the sum to the spins within a disc of radius $\rho$.  A numerical summation\cite{Archambault1} shows that
\begin{equation}
\sum_{r=1}^N G^2(r) = \frac{1}{N} \sum_{q\ne 0} (\frac{1}{\epsilon_q})^2 = \frac{N}{c},
\end{equation}
with $c=258.59...$.  Letting
\begin{equation}
u = \frac{\pi^2}{8N} \frac{1}{\delta^2(\ln{2\rho})}\exp{(-\frac{\pi G(0)}{2\delta(\ln{2\rho})})},
\end{equation}
\begin{equation}
du = \frac{\pi^2}{8N}(\frac{\pi G(0)}{2\delta}-2) \frac{1}{\delta^3} \exp{(-\frac{\pi G(0)}{2\delta})} d\delta.
\end{equation}
Collecting these together, 
\begin{equation}
\label{spinpartway}
\begin{split}
&\frac{\chi_S T}{N} =  \frac{\pi^2 \aleph(\rho)}{8cN} \frac{1}{\delta^2(\ln{2\rho})}\exp{(-\frac{\pi G(0)}{2\delta(\ln{2\rho})})} \big{|}^{\Delta/2}_{1/\sqrt{\pi}} \\
& - \frac{\pi^2}{8c}\int^{\delta(\ln{\Delta})}_{\delta(\ln{\frac{2}{\sqrt{\pi}}})} d\delta \; (\frac{2\rho(\delta)}{\Delta})^2 (\frac{\pi G(0)}{2\delta}-2) \frac{1}{\delta^3} \exp{(-\frac{\pi G(0)}{2\delta})}.
\end{split}
\end{equation}
Since $\aleph(\rho=\Delta/2)=N$, the lower limit in the first line of eq.(\ref{spinpartway}) is $\sim 1/N$ smaller than the upper limit, and is neglected.  In the remaining integral, the expression $2\rho(\delta,\omega)$ can be evaluated by isolating the term $\ln(L)=\ln({2\rho)}$ in the geometric scaling equations (\ref{exactd}) or (\ref{smalld}) and (\ref{L(w)}) that give $\delta(\ln{2\rho}, \omega)$.
\begin{equation}
\begin{split}
&\frac{\chi_S T}{N} =  \frac{\pi^2}{8c} \frac{1}{\delta^2(\ln{\Delta})}\exp{(-\frac{\pi G(0)}{2\delta(\ln{\Delta})})} \\
&- \frac{\pi^2}{8c}\int^{\delta(\ln{\Delta})}_{\delta(\ln{\frac{2}{\sqrt{\pi}}})} d\delta \; (\frac{2\rho(\delta)}{\Delta})^2 (\frac{\pi G(0)}{2\delta}-2) \frac{1}{\delta^3} \exp{(-\frac{\pi G(0)}{2\delta})}.
\end{split}
\end{equation}
Again, in the low temperature limit, $\delta=\pi J/2k_BT$ scales very slowly with size, so that the coupling $J$ is essentially constant.  This means the integral portion of the spin wave susceptibility is zero because the limits of the integral are essentially the same.  The first term is then equivalent to
\begin{equation}
\frac{\chi_S T}{N} = \frac{T^2}{2J^2 c} \langle M \rangle ^2,
\end{equation}
in agreement with Archambault \textit{et al.} \cite{Archambault1}

\begin{acknowledgments}
Financial support for this work  was provided by the Natural Sciences and Engineering Research Council of Canada (NSERC) through the Discovery program.  I am indebted to Sung-Sik Lee of McMaster University for many useful discussions, and thank the referees for constructive comments.  \end{acknowledgments}

\bibliography{chi_RG_expt.bib}

\end{document}